\documentclass[aps,superscriptaddress,pra,twocolumn,showpacs,letterpaper,tighten,float]{revtex4-1}
\usepackage{amssymb}
\usepackage{amsbsy}
\usepackage{amsmath}
\usepackage{epsfig}
\usepackage{graphicx}
\usepackage{array}
\usepackage[utf8]{inputenc}
\usepackage{textcomp}
\usepackage{color}
\usepackage{bm}

\DeclareMathOperator{\sgn}{sgn}

\definecolor{myblue}{rgb}{.93, .93, 1}
\newcommand*\mybluebox[1]{%
\colorbox{myblue}{\hspace{1em}#1\hspace{1em}}}
\newcommand*\myFbluebox[1]{%
\fcolorbox{black}{myblue}{\hspace{1em}#1\hspace{1em}}}

\newcommand{\bsub}{\begin{subequations}}
\newcommand{\esub}{\end{subequations}}
\newcommand{\beq}{\begin{empheq}[box=\mybluebox]{align}}
\newcommand{\beqF}{\begin{empheq}[box=\myFbluebox]{align}}

		\newcommand{\ord}[1]{\bm{\mathit{O}}\left(#1\right)}

		\newcommand{\vex}[1]{\bm{\mathrm{#1}}}

		\newcommand{\ket}[1]{\left| {#1} \right\rangle}

		\newcommand{\tauel}{\tau_{\mathsf{el}}}
		\newcommand{\e}{\varepsilon}

\begin{document}

\title{Twisting Anderson pseudospins with light:\\ Quench dynamics in THz-pumped BCS superconductors}

\author{Yang-Zhi~Chou} \email{YangZhi.Chou@colorado.edu}
\affiliation{Department of Physics and Astronomy, Rice University, Houston, Texas 77005, USA}
\affiliation{Department of Physics and Center for Theoretical Quantum Matter, University of Colorado Boulder, Boulder, Colorado 80309, USA}
\author{Yunxiang~Liao}
\affiliation{Department of Physics and Astronomy, Rice University, Houston, Texas 77005, USA}
\author{Matthew~S.~Foster}
\affiliation{Department of Physics and Astronomy, Rice University, Houston, Texas 77005, USA}
\affiliation{Rice Center for Quantum Materials, Rice University, Houston, Texas 77005, USA}

\date{\today}

\pacs{74.40.Gh,78.47.J-,03.75.Kk}

\begin{abstract}
We study the preparation (pump) and the detection (probe) of 
far-from-equilibrium BCS superconductor dynamics in THz pump-probe experiments.
In a recent experiment
[R. Matsunaga, Y. I. Hamada, K. Makise, Y. Uzawa, H. Terai, Z. Wang, and R. Shimano, Phys. Rev. Lett. {\bf 111}, 057002 (2013)],
an intense monocycle THz pulse with center frequency $\omega \simeq \Delta$ was injected into a superconductor with
BCS gap $\Delta$; the subsequent post-pump evolution was detected via the optical conductivity.
It was argued that nonlinear coupling of the pump to the Anderson pseudospins of the superconductor 
induces coherent dynamics of the Higgs (amplitude) mode $\Delta(t)$. We validate this picture in a
two-dimensional 
BCS model 
with a combination of exact numerics and 
the Lax reduction method, and we compute the nonequilibrium phase diagram as a function of the pump intensity.
The main effect of the pump is to scramble the orientations of Anderson pseudospins along the Fermi surface
by twisting them in the $xy$-plane. 
We show that more intense pump pulses can induce a far-from-equilibrium phase of gapless superconductivity (``phase I''),
originally predicted in the context of interaction quenches in ultracold atoms. 
We show that the THz pump method can reach phase I at much lower energy densities than an interaction
quench, and we demonstrate that Lax reduction (tied to the integrability of the BCS Hamiltonian) provides
a general quantitative tool for computing coherent BCS dynamics. We also calculate the Mattis-Bardeen 
optical conductivity for the nonequilibrium states discussed here.
\end{abstract}

\maketitle

\tableofcontents

\section{Introduction}

A series of recent optical pump-probe experiments has reignited interest in far-from-equilibrium superconductivity
\cite{Fausti2011,Matsunaga2012,Matsunaga2013,Mansart2013,Matsunaga2014,Hu2014,Mankowsky2015,Matsunaga2015,Mitrano2016,Nicoletti2016}. 
Most of the experiments fall into two broad classes. 
In the first class, mid-infrared radiation ($\sim$ 10 THz) is injected with the goal of enhancing pairing, 
either by destabilizing competing orders \cite{Fausti2011} 
or by exciting optical phonons that participate in pairing \cite{Hu2014,Mankowsky2015,Mitrano2016}. 
The essential idea of the latter is that by modulating the ``pairing glue,'' 
one might induce transient high(er) temperature superconductivity than is possible in equilibrium \cite{Knap2015,Kennes2016}.

A complication of any high frequency excitation with $\omega \gg \Delta$ ($\Delta$ is the BCS gap) 
is that the radiation can break Cooper pairs into hot quasiparticles, and these can serve as an efficient mechanism for 
rapid dissipation and thermalization. The second class of experiments \cite{Matsunaga2012,Matsunaga2013,Matsunaga2014}
used lower frequency radiation ($\sim$ 1 THz), which might mitigate heating. In Ref.~\cite{Matsunaga2013},
a near monocycle pulse with center frequency $\omega \simeq \Delta$ was injected into a low-temperature BCS
superconductor. Because most of the spectral weight lies below the optical gap edge $2 \Delta$,
a weak pulse would not be expected to couple strongly to the sample. 
However, an intense pulse as used in Ref.~\cite{Matsunaga2013}
couples \emph{nonlinearly} \cite{Tsuji2015}
to the Cooper pairs (``Anderson pseudospins'' \cite{Anderson1958}) of the superconductor,
and it was argued that this leads to a coherent excitation of the Higgs
amplitude mode $\Delta(t)$ \cite{Volkov1974,LittleWood1981,Barankov2004Collective_Rabi,Yuzbashyan2005PRB,Yuzbashyan2005JPA,Barankov2006PRA,Yuzbashyan2006Relaxation,Barankov2006,Yuzbashyan2006Dynamical_Vanishing,Gurarie2009,PapenKort2007,PapenKort2008,Krull2014,Tsuji2015,Yuzbashyan2015,Cea2016,Varma2002,Pekker2015}. 
The advent of ultrafast pump-probe detection can allow the observation of 
coherent many-body quantum dynamics in the solid state within a finite temporal window
(``prethermalization plateau'' \cite{Kinoshita2006,Rigol2007}). 
In the experiments \cite{Matsunaga2012,Matsunaga2013,Matsunaga2014} detection is performed over a window
of about 10 picoseconds (ps), well before thermalization occurs (likely due to acoustic phonons on a timescale of 100 ps \cite{Lee2003}).

\begin{figure}[t!]
\includegraphics[width=0.45\textwidth]{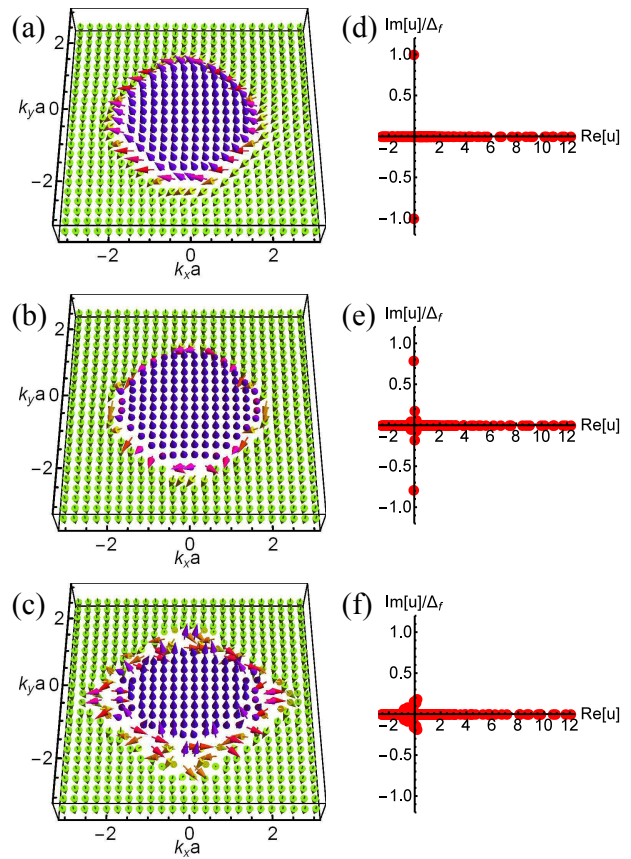}
\caption{The Anderson pseudospin textures and the Lax roots of the spectral polynomial immediately after 
the application of an intense THz pulse. 
(a)-(c) The spin textures after light exposure, corresponding to the
roots shown in (d)-(f), respectively.
The pair of roots away from the real axis is called isolated and these roots encode the key properties 
of the BCS state.  
The system is a quarter-filled square lattice tight-binding model with 24-by-24 sites
(note that our subsequent quantitative results are obtained from much larger systems).
$\Delta_f = 0.2J$ is the ground state order parameter of the BCS Hamiltonian; $J$ is the tight-binding parameter. 
The pump energies [$\tilde{A}^2$ defined in Eq.~(\ref{tilA2Def})] are (a),(d)
$\tilde{A}^2=0.1\Delta_f$ (very weak deformation from the BCS ground state), 
(b),(e), $\tilde{A}^2=2.5\Delta_f$ (intermediate strength deformation), 
and (c),(f)
$\tilde{A}^2=10\Delta_f$ (strong deformation). 
The isolated roots $u^{\pm}_0\approx\pm i\Delta_{\infty}$
for the deformed spin textures encode
the asymptotic value of $\Delta(t \!\! \rightarrow\!\! \infty) = \Delta_\infty$ in the 
``pre-thermalization plateau.'' 
(The chemical potential is absorbed into the dispersion in these plots.)
More intense pulses push the isolated roots towards the real axis.
For strong pulses, the isolated roots merge with the real axis and the 
system enters a phase of dynamical gapless
superconductivity \cite{Barankov2006,Yuzbashyan2006Dynamical_Vanishing} with $\Delta_\infty = 0$; 
see the ``quench phase diagram'' in Fig.~\ref{Fig:Phase_Diagram}.
}
\label{Fig:Spin_and_Lax}
\end{figure}

In this paper, we model pump-probe experiments as typified by Ref.~\cite{Matsunaga2013},
using a combination of numerics and the Lax reduction method \cite{Yuzbashyan2006Relaxation}. 
We view the THz pump excitation as the preparation of an initial condition to the subsequent
free (unperturbed) BCS time evolution, i.e.\ as a type of quantum quench 
\cite{Greiner2002,Barankov2004Collective_Rabi,Barankov2006,Yuzbashyan2005PRB,Yuzbashyan2005JPA,Yuzbashyan2005PRB_FBcond,Yuzbashyan2006Relaxation,Barankov2006PRA,Calabrese2006,Gurarie2009}.
We show that the main effect of the pump is to scramble the orientations of Anderson pseudospins along the 
Fermi surface by twisting them in the $xy$-plane.
We also show that more intense pump pulses can induce a far-from-equilibrium phase of gapless superconductivity (``phase I''),
originally predicted in the context of interaction quenches in ultracold atomic systems \cite{Barankov2006,Yuzbashyan2006Dynamical_Vanishing}.
An intense THz pump can reach phase I at much lower energy densities than an interaction
quench.

Most existing theoretical studies of coherent far-from-equilibrium BCS superfluid
dynamics
focus on interaction quenches of isolated systems, 
as could be realized in ultracold atom experiments \cite{BlochRMP2008,PolkovnikovRMP2011,Eisert2015}. 
Due mainly to the integrability of the BCS model \cite{Yuzbashyan2005PRB,Yuzbashyan2005JPA,Yuzbashyan2006Relaxation}, 
many asymptotically exact results are already known.
The most important is the identification of 
three different nonequilibrium phases (dubbed ``I, II, and III'' in \cite{Foster2013,Yuzbashyan2015}, reviewed below),
predicted to occur in quenches of 
an $s$-wave superconductor 
\cite{Barankov2006,Yuzbashyan2006Dynamical_Vanishing}, 
$p+ip$ superfluid 
\cite{Foster2013,Foster2014,Liao2015}, 
BCS-BEC condensate \cite{Yuzbashyan2005PRB_FBcond,Gurarie2009,Yuzbashyan2015}, 
and spin-orbit-coupled fermion condensate 
\cite{Dong2015,Dzero2015}.

Ultrafast pump-probe spectra in BCS superconductors have been studied numerically using the density matrix formalism 
\cite{PapenKort2007,PapenKort2008,Krull2014}.
These studies revealed asymptotic post-pump dynamics identical to a small
``phase II'' interaction quench, in which $\Delta(t)$ approaches a nonequilibrium value $\Delta_\infty$ as $t \rightarrow \infty$
(neglecting dissipative processes that would ultimately induce thermalization).
The approach to $\Delta_\infty$ involves a characteristic damped oscillation at frequency $2 \Delta_\infty$ \cite{Volkov1974,Yuzbashyan2006Relaxation},
also seen in the experiment \cite{Matsunaga2013}. 
The calculations in Refs.~\cite{PapenKort2007,PapenKort2008,Krull2014}
assumed coupling to the electromagnetic field
due to the 
finite (but large) photon wavelength $\lambda \gtrsim 300$ $\mu$m. 
In the present paper, 
we 
classify the dynamics according to the 
interaction quench phase diagram \cite{Barankov2006,Yuzbashyan2006Dynamical_Vanishing,Foster2013,Yuzbashyan2015},
and 
we
assume that the strongest coupling to the pump excitation is due to non-quadratic band curvature \cite{Tsuji2015}.  
A related experiment \cite{Matsunaga2014} shows third harmonic generation in the steady-state response;
this was studied theoretically in \cite{Tsuji2015,Cea2016}.

In this work, we demonstrate that the Lax reduction method \cite{Yuzbashyan2006Relaxation} 
(tied to the integrability of the BCS Hamiltonian) provides a general quantitative tool for computing 
coherent nonlinear BCS dynamics.
We also calculate the Mattis-Bardeen 
\cite{Mattis1958,Tinkham1996}
optical conductivity for the nonequilibrium superconducting states predicted here.

The rest of this paper is organized as follows: 
In Sec.~\ref{Sec:Results}, we review the interaction quench dynamics of $s$-wave BCS superconductors.
We then highlight our main findings for a THz pump probe, including the dynamical phase diagram given by
Fig.~\ref{Fig:Phase_Diagram}
and predictions for the optical conductivity.
In Sec.~\ref{Sec:pump}, we provide details of the light-superconductor interaction that twists 
Anderson pseudospins during the application of the pump.
In Sec.~\ref{Sec:Probe}, we calculate the optical conductivity for nonequilibrium superconductors. 
In Sec.~\ref{Sec:Thermalization}, we introduce the ``R-ratio'' to quantify the relative energy
injected by the pump, and we compare the threshold energies for reaching phase I in THz pump versus
interaction quenches. We summarize in Sec.~\ref{Sec:Discussion}.


\section{Quantum quench via pump-probe: Main results \label{Sec:Results}}

In this section, we review the dynamical phases of nonequilibrium BCS superconductors 
induced by interaction quenches. We then summarize our main results for a THz pump probe. 
Details of our calculations appear in Secs.~\ref{Sec:pump} and \ref{Sec:Probe}.

\subsection{BCS model, review of interaction-quench dynamics, and relation to the ``THz pump quench'' \label{Sec: model, intq review}}

The BCS Hamiltonian can be expressed in terms of 
Anderson pseudospins \cite{Anderson1958},
\begin{align}\label{Eq:And_spin_H}
	H_{\text{BCS}}
	=
	\sum_{\vex{k}}
	2\tilde{\varepsilon}_{\vex{k} + \frac{e}{c} \vex{A}(t)}\,s^{z}_{\vex{k}}
	-
	G
	\sum_{\vex{k},\vex{k}'}s_{\vex{k}}^+ s_{\vex{k}'}^-,
\end{align}
where
$\tilde{\varepsilon}_{\vex{k}}=\varepsilon_{\vex{k}}-\mu$, 
$\varepsilon_{\vex{k}}$ is the single particle dispersion, 
$\mu$ is the chemical potential,
and $G$ is the coupling strength. 
The spin operators are
\begin{align}
	s^z_{\vex{k}}
	\equiv
	{\textstyle{\frac{1}{2}}}
	\left(c^{\dagger}_{\vex{k}\uparrow}c_{\vex{k}\uparrow}+c^{\dagger}_{-\vex{k}\downarrow}c_{-\vex{k}\downarrow}-1\right),
	\quad
	s^+_{\vex{k}}
	\equiv 
	c^{\dagger}_{\vex{k}\uparrow}c^{\dagger}_{-\vex{k}\downarrow},
\end{align}
and $s^-_{\vex{k}} = \left(s^+_{\vex{k}}\right)^\dagger$.
In the thermodynamic limit, Eq.~(\ref{Eq:And_spin_H}) is equivalent to the mean field Hamiltonian,
\bsub\label{Eq:BCS_MFT}
\begin{align}
	\label{Eq:MF_H}
	H_{\text{BCS}}
	&\rightarrow 
	H_{\text{MF}}
	=
	-\sum_{\vex{k}}\vec{s}_{\vex{k}}\cdot\vec{B}_{\vex{k}},
\\
	\label{Eq:B_field}
	\vec{B}_{\vex{k}}
	&=
	-2\left\{
		\left[{\varepsilon}_{\vex{k} + \frac{e}{c} \vex{A}(t)}-\mu\right]\hat{z}
		+ 
		\Delta_x \, \hat{x} 
		+ 
		\Delta_y \, \hat{y}
	\right\}.
\end{align}
\esub
The instantaneous order parameter is self-consistently determined
\begin{align}
	\label{Eq:GapEqn}
	\Delta(t) \equiv \Delta_x - i \Delta_y = -G \sum_{\vex{k}}\langle s^-_{\vex{k}}(t)\rangle.
\end{align}
The equation of motion for an Anderson pseudospin is
\begin{align}\label{SpinEOM}
	\dot{\vec{s}}_{\vex{k}}=-\vec{B}_{\vex{k}}\times\vec{s}_{\vex{k}}.
\end{align}
The vector potential $\vex{A}(t)$ will be used to encode the pump electric field \cite{Tsuji2015}. 

Previous studies of quench dynamics in superconductors focused on interaction quenches, in which
the ground state of Eq.~(\ref{Eq:And_spin_H}) with an initial coupling strength $G_i$ is time-evolved according
to the post quench Hamiltonian, given by Eq.~(\ref{Eq:And_spin_H}), with ``final'' coupling strength $G_f$ 
(and $\vex{A} = 0$ at all times).
The long-time behavior of the order parameter $\Delta(t)$ following the quench 
can be predicted with the Lax reduction method \cite{Yuzbashyan2006Relaxation}, which
exploits the integrability of $H_{\text{BCS}}$.

Interaction quenches from an initial BCS state can produce three possible nonequilibrium phases 
\cite{Yuzbashyan2006Relaxation,Barankov2006,Yuzbashyan2006Dynamical_Vanishing};
an interaction-quench dynamical phase diagram can be constructed in the $\Delta_i$-$\Delta_f$ plane,
where $\Delta_i$ ($\Delta_f$) denotes the ground state BCS gap of the pre-quench (post-quench) Hamiltonian,
and $\Delta_i = \Delta_f$ corresponds to equilibrium (no quench) \cite{Foster2013,Yuzbashyan2015}.  
Small quenches with $\Delta_i \sim \Delta_f$ reside in phase II, wherein $\Delta(t)$ asymptotes 
to a non-zero constant value $\Delta_\infty$ as $t \rightarrow \infty$. Note
that $\Delta_i \neq \Delta_f \neq \Delta_\infty$ for a quench.
Large quenches from strong to weak pairing ($\Delta_f \ll \Delta_i$) can induce phase I,
in which $\Delta(t) \rightarrow 0$ \cite{Yuzbashyan2006Dynamical_Vanishing,Barankov2006};
this is a phase of gapless (fluctuating) superconductivity. 
Large quenches from weak to strong pairing ($\Delta_i \ll \Delta_f$) can induce phase III,
in which $\Delta(t)$ exhibits persistent oscillations \cite{Barankov2004Collective_Rabi,Barankov2006}. 
Phase III is a self-generated Floquet phase without external periodic driving \cite{Foster2014,Liao2015},
and is connected to the instability of the normal state.

In the THz pump-probe experiments of Refs.~\cite{Matsunaga2013,Matsunaga2015}, the injected
pump pulse can be viewed as preparing an initial condition for subsequent free evolution
in the absence of the field. In this picture, the interaction strength $G$ is not modified
by the radiation. Instead, for a system at zero temperature, the pump pulse 
deforms the BCS ground state into a new pure state over the duration of the pulse [wherein $\vex{A}(t) \neq 0$]. 
The deformed state subsequently evolves with the {\it original} Hamiltonian and $\vex{A}(t) = 0$. 
In analogy with the interaction quench, we identify the original Hamiltonian as the post-quench Hamiltonian, 
i.e., the ground state gap is denoted $\Delta_f$ in the following.

\begin{figure}
	\centering
\includegraphics[width=0.4\textwidth]{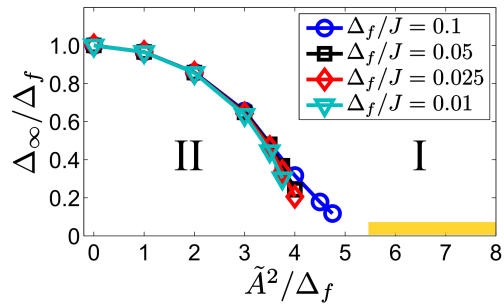}
\caption{The nonequilibrium phase diagram of an $s$-wave BCS superconductor 
subject to an ultrashort ($\simeq$ monocycle) THz pulse with center frequency $\omega \simeq \Delta_f$ (``pump quench''). 
We consider Eq.~(\ref{Eq:And_spin_H}) on a 2D square lattice with a linear size $L=1000$ 
(1000$\times$1000 sites)
at quarter filling. 
Values of $\Delta_{\infty}$ are extracted from the isolated Lax roots corresponding 
to the spin configuration immediately after cessation of the pump pulse. 
We verified that these results agree with 
$\Delta(t \rightarrow \infty)$ extracted from Eq.~(\ref{Eq:GapEqn}), using the
numerical integration of Eq.~(\ref{SpinEOM}) to large times. 
We plot $\Delta_{\infty}$ as a function of the peak relative field intensity $\tilde{A}^2$ [Eq.~(\ref{tilA2Def})] 
with different values of $\Delta_f/J$. 
Here $\Delta_f$ denotes the ground state BCS gap, and $J$ is the hopping strength.
The polarization angle $\alpha=0$ for all cases, i.e.,
the radiation is polarized along the $x$-axis. 
The results weakly depend on $\Delta_f/J$.  
More intense pulses produce larger deformations of the pseudospins along the Fermi surface 
(see Fig.~\ref{Fig:Spin_and_Lax}), leading to a suppression of the asymptotic BCS gap $\Delta_\infty$. 
For small values of $\Delta_{\infty}$, finite size effects become significant, and
it becomes difficult to resolve the isolated roots via the Lax method. For these
cases the results were obtained via numerical integration of Eq.~(\ref{SpinEOM}). 
Phase I ($\Delta_\infty = 0$) can be achieved with $\tilde{A}^2/\Delta_f>5.5$ (yellow shaded region) in all cases. 
The phase boundary for $\Delta_f/J=0.01$ is close to $\tilde{A}^2/\Delta_f=4.5$.
}
\label{Fig:Phase_Diagram}
\end{figure}

\begin{figure}[t]
	\centering
	\includegraphics[width=0.4\textwidth]{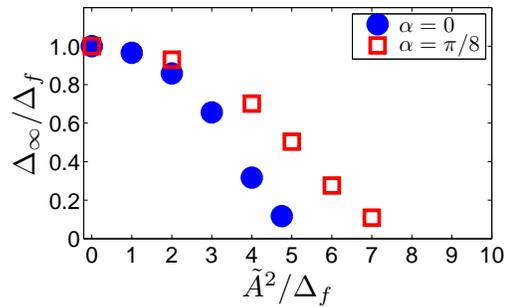}
	\caption{The polarization angle dependence of $\Delta_\infty$ for the pump quench. 
	The system is the same as in Fig.~\ref{Fig:Phase_Diagram}. 
	We plot $\Delta_{\infty}$ as a function of pump energy ($\tilde{A}^2$) with two polarization angles, 
	$\alpha=0,\pi/8$. Here we set $\Delta_f/J=0.1$. 
	Radiation polarized along the $x$-axis ($\alpha=0$) 
	generates nonequilibrium dynamics most efficiently, 
	due to the density of states accumulation along the parallel Fermi line 
	segments with $k_x = 0$. 
	For $\alpha = \pi/4$, we did not observe phase I dynamics.
	To determine whether the latter is possible, 
	it would be necessary to retain higher order terms $\ord{\tilde{A}^4}$ 
	in Eqs.~(\ref{Eq:dispersion_and_A_SL}) and (\ref{Eq:F_k}). 
	}
	\label{Fig:Order_parameter_Lax1000}
\end{figure}

\begin{figure}[b]
	\centering
	\includegraphics[width=0.4\textwidth]{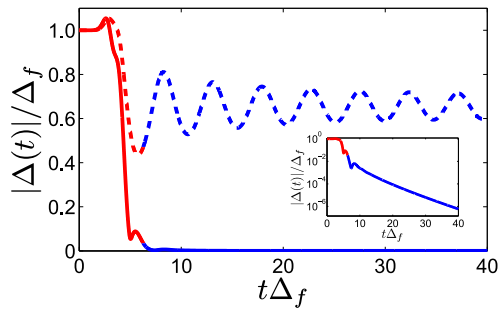}
	\caption{The real time evolution of the superconducting order parameter. 
	The system is a square lattice with linear size $L=2000$ at quarter filling. 
	The pump-pulse polarization angle $\alpha$ is set to 0, and $\Delta_f/J=0.1$. 
	The dashed line corresponds to pump energy $\tilde{A}^2=3\Delta_f$ (phase II); 
	the solid line corresponds to pump energy $\tilde{A}^2=5.5\Delta_f$ (phase I). 
	The red regions indicate the evolution during the application of the 
	pump pulse. 
	The blue regions mark the subsequent free time evolution with the time-independent BCS Hamiltonian. 
	Inset: The phase I dynamics of $\tilde{A}^2=5.5\Delta_f$. The order parameter decays exponentially in time \cite{Yuzbashyan2006Dynamical_Vanishing}.
	}
	\label{Fig:Order_parameter_L2000_alpha0}
\end{figure}

\begin{figure}
	\centering
	\includegraphics[width=0.4\textwidth]{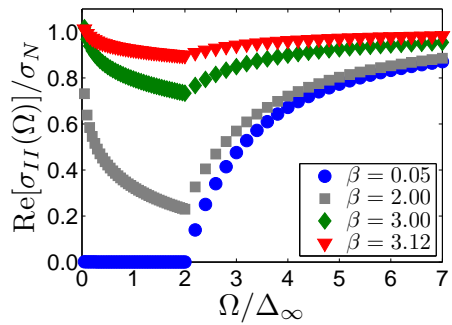}
	\caption{The real part of the Mattis-Bardeen conductivity in phase II.
	The result depends on $\Delta_\infty$, but also on the full nonequilibrium distribution function
	$n_{\vex{k}}$ that determines the occupation of quasiparticle states in the asymptotic steady state. 
	For this plot and Fig.~\ref{Fig:MB_ImPart_phaseII}, we use the distribution function for
	an interaction quench whose analytical form is exactly known. We expect that results for
	the pump quench are qualitatively identical. 
	The interaction quench is specified by the quench parameter 
	$
	\beta
	\equiv
	2\left(
		\frac{1}{G_f}
		-
		\frac{1}{G_i}
	\right)
	$, 
	where $G_i$ ($G_f$) is the interaction strength of the pre-quench (post-quench) Hamiltonian. 
	Since the Mattis-Bardeen formulae \cite{Mattis1958} assume the dirty limit $\Delta \tauel \ll 1$ ($\tauel$ is the elastic scattering time),
	we consider a 2D particle-hole symmetric superconductor with a uniform density of states. 
	The dynamical phase II-I boundary is located at $\beta=\beta_c=\pi$. 
	The real part of conductivity shows a cusp at $\Omega=2\Delta_{\infty}$ for all phase II quenches. 
	Although our system is specified at all times by a nonequilibrium BCS pure state, the 
	optical conductivity for a quench ``looks'' thermal.
	In particular, as $\beta$ approaches the phase II-I boundary $\beta_c=\pi$, the real part
	of the optical conductivity behaves the same way as an equilibrium superconductor 
	at temperature $T$
	approaching
	$T_c$ from below \cite{Tinkham1996}. The steady-state optical conductivity for phase I is indistinguishable from
	a normal metal.
	}
	\label{Fig:MB_RePart_phaseII}
\end{figure}

\begin{figure}[b]
	\centering
	\includegraphics[width=0.4\textwidth]{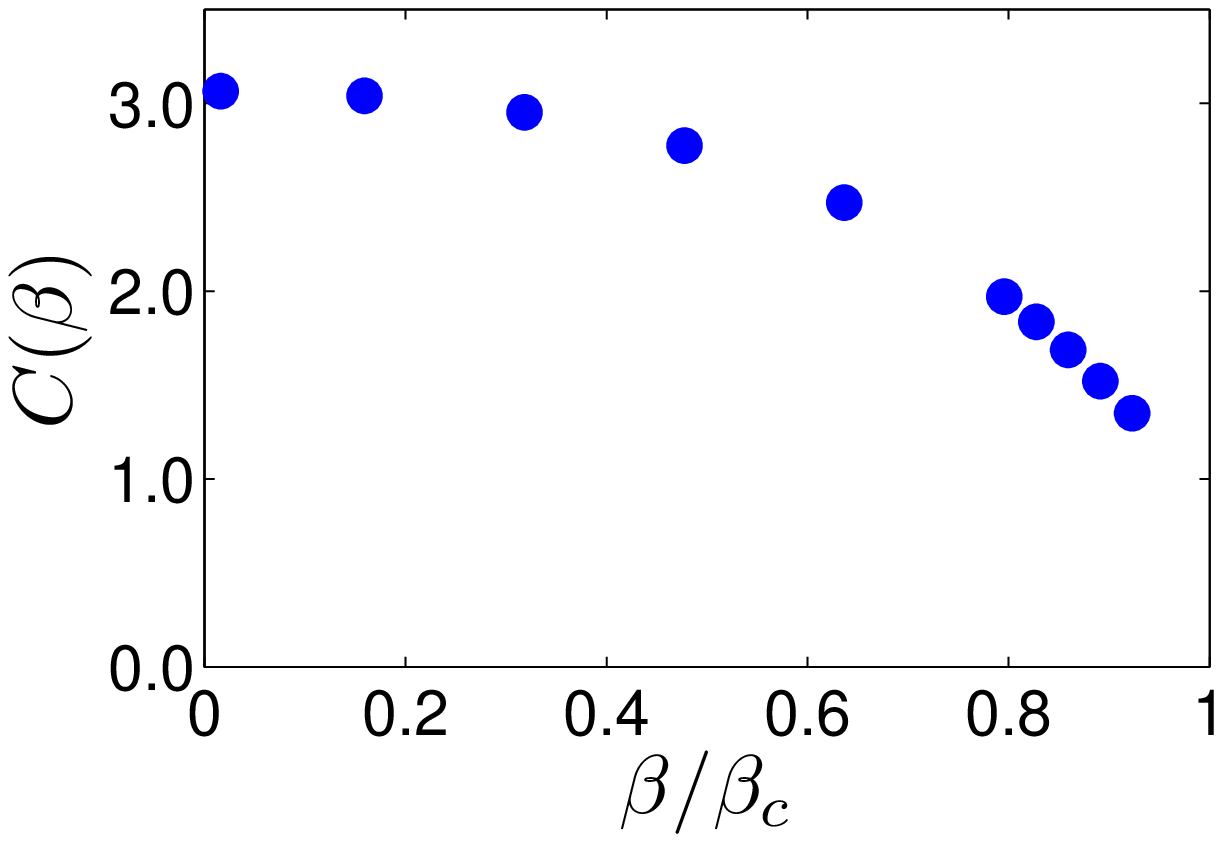}
	\caption{The prefactor of the imaginary part of the Mattis-Bardeen conductivity, 
	corresponding to the phase II quenches shown in Fig.~\ref{Fig:MB_RePart_phaseII}.
	The horizontal axis specifies the quench parameter 
	$
	\beta
	=
	2\left(
		\frac{1}{G_f}
		-
		\frac{1}{G_i}
	\right)
	$. 
	The imaginary part of the conductivity can be fitted by 
	$
	\text{Im}[\sigma_{\text{II}}(\Omega)]/\sigma_N
	= 
	C(\beta)
	\,
	\Delta_{\infty}
	/
	\Omega
	$ 
	for 
	$\Omega \ll 2\Delta_{\infty}$. The value in equilibrium with $T \ll T_c$ is $C(0) = \pi$. 
	$\beta_c=\pi$ is the critical value of the quench parameter that separates phase II and phase I.}
	\label{Fig:MB_ImPart_phaseII}
\end{figure}

\subsection{Pump-pulse quench: Results and phase diagram}

Light couples via the vector potential to the $z$-component of the Anderson pseudospin $B$-field [Eq.~(\ref{Eq:B_field})]. 
We neglect the small photon momentum in the following, and encode the uniform electric field of the THz pump via
$\vex{E}(t) = - (1/c) \, d \vex{A}(t) / d t$. 
The strength of the pump excitation can be measured by the peak field intensity (energy)
\begin{align}\label{tilA2Def}
	\tilde{A}^2
	\equiv
	J \, a^2\frac{e^2}{c^2} \max \vex{A}^2,
\end{align}
where $a$ denotes the lattice spacing and $J$ is the hopping strength that sets
the electronic bandwidth. 
For pump energies $\tilde{A}^2 \ll J$, the
dispersion 
$
	\varepsilon_{\vex{k} + e \vex{A}(t) / c}
$ 
can be expanded in powers of $A$. 
In a clean single band superconductor, linear coupling to $\vex{A}(t)$ does not induce
time evolution from a pure BCS state \cite{MahanBook,Dai2016}; 
the same is true for optical subgap excitation at
frequency $\omega < 2 \Delta$ in 
a dirty superconductor. An intense pump pulse can couple through the band curvature 
at order $A^2$ as long as the dispersion is not purely parabolic
[see Eqs.~(\ref{Eq:B_field_Light_SC}) and (\ref{Eq:dispersion_and_A})]. 

We consider a two-dimensional (2D) square lattice model at quarter-filling \cite{quarter_filling}.
Application of an intense, monocycle pump pulse with center frequency $\omega \simeq \Delta_f$
($\Delta_f$ denotes the ground state gap) generates dynamics that deform
the Anderson pseudospin texture of the initial state. Within the self-consistent
mean field framework given by Eq.~(\ref{Eq:BCS_MFT}), the light-matter interaction does
not break Cooper pairs (remove Anderson pseudospins), and the system resides
in a BCS product state at all times (albeit one with time-evolving coherence factors).
In a real experiment such as Ref.~\cite{Matsunaga2013}, the superconductor is disordered
and quasiparticles are always excited by the linear coupling to $\vex{A}(t)$,
since some spectral weight of the ultrashort pump pulse extends above $\omega = 2 \Delta_f$. 
We assume that the effects of these quasiparticles on the coherent evolution of 
$\Delta(t)$ can be neglected on the timescale of the experiment, which is typically
10 ps \cite{Matsunaga2012,Matsunaga2013,Matsunaga2014,Matsunaga2015}.  

In the presence of the pump-pulse light, we use the fourth order Runge-Kutta method to 
numerically integrate the spin equations of motion, given by Eq.~(\ref{SpinEOM}). 
After cessation of the pump, we extract the spin configuration and construct the 
corresponding spectral polynomial via the Lax vector \cite{Yuzbashyan2005PRB,Yuzbashyan2005JPA} (also see Appendix~\ref{Sec:App:Lax}).
The spectral polynomial is a conserved quantity under time evolution with the 
[unperturbed, $\vex{A}(t) = 0$] BCS Hamiltonian; its roots (``Lax roots'') are therefore
also conserved. The isolated Lax roots occur in complex conjugate pairs, and 
encode the nature of the nonequilibrium state, and
this is a robust, \emph{topological} classification 
scheme \cite{Yuzbashyan2006Relaxation}. 
In particular, interaction quenches that produce phases I, II, and III (described above) respectively correspond
to spectral polynomials with zero, one, or two pairs of isolated roots \cite{Yuzbashyan2006Dynamical_Vanishing,Barankov2006}.
This scheme also applies to topological $p$-wave superfluids \cite{Foster2013} 
and strongly paired BCS-BEC $s$-wave fermionic condensates \cite{Yuzbashyan2015}.   
The isolated roots exactly determine $\Delta(t)$ in the $t \rightarrow \infty$ limit. 
For phases II and I, $\Delta(t \rightarrow \infty) = \Delta_\infty$, 
where $\Delta_\infty$ is a nonzero constant (zero) in phase II (I). 
In phase II, $\Delta_\infty$ is equal to the modulus of the imaginary part of the single isolated root pair. 
 
For the pump-pulse quenches, we found that the spectral polynomial evaluated in terms
of the post-pump Anderson pseudospin texture exhibits one or zero pairs of isolated roots, 
depending upon the degree of deformation from the initial BCS ground state. 
These results are consistent with the real time numerical simulations.
The spin textures and the roots of the spectral polynomial are demonstrated for small
system sizes in Fig~\ref{Fig:Spin_and_Lax}. 
A quench phase diagram is constructed using much larger systems in Fig.~\ref{Fig:Phase_Diagram}. 
In phase II, $\Delta_{\infty}$ only weakly depends on the value of $\Delta_f/J$ 
(see Fig~\ref{Fig:Phase_Diagram}); 
here $\Delta_f$ is the order parameter of the ground state, and 
$J$ is the tight-binding parameter of the square lattice. 
This result suggests that Eq.~(\ref{Eq:dispersion_and_A}) is a reasonable approximation 
for capturing the nonlinear light-superconductor coupling. 
In Fig.~\ref{Fig:Order_parameter_Lax1000}, the dependence of $\Delta_{\infty}$ 
on the polarization angle of the pump-pulse electric field is shown. 

We have confirmed the isolated root predictions for phase II numerically up to $L=4000$ 
with fourth order Runge-Kutta dynamics. ($L$ is the linear size of the square lattice.) 
For sufficiently large pump energies, 
we find $\Delta_\infty \rightarrow 0$, consistent with phase I dynamics. 
The time evolution of the order parameter amplitudes in phases II and I is shown in 
Fig.~\ref{Fig:Order_parameter_L2000_alpha0}.

\subsection{Optical conductivity in phases I and II \label{Sec: OptI,II}}

Besides the mechanism of the pump pulse quench, 
we have investigated the optical conductivity in phases II and I.
The steady-state optical conductivity in the prethermalization plateau 
can be determined by the absorption or reflection of the probe pulse. 
Real low-temperature superconductors typically reside in the dirty limit,
wherein $\Delta \tauel \ll 1$. Here $\tauel$ is the lifetime due to elastic
impurity scattering. We reformulate the Mattis-Bardeen formulae \cite{Mattis1958}
for the real and imaginary parts of the optical conductivity in dirty superconductors,
adapting them to steady-state nonequilibrium superconductivity. 

For phases II and I, the generalized Mattis-Bardeen formulae depend on $\Delta_\infty$,
but also on the full nonequilibrium quasiparticle distribution function $n_k$. 
Here we employ the exact result for $n_k$ previously determined 
\cite{Yuzbashyan2006Dynamical_Vanishing}
for 
interaction quenches. The optical conductivity for the pump-probe quench will
be qualitatively the same within each dynamical phase.   

For a small phase II quench, the order parameter approaches to $\Delta_{\infty} \neq 0$ in the long time limit. 
The real part of the optical conductivity exhibits a cusp at frequency $\Omega=2\Delta_{\infty}$, 
but spectral weight extends down to zero (filling the gap) frequency for any nonzero quench. 
This is very similar to an equilibrium superconductor at finite temperature $T < T_c$,
despite the fact that our system is described by a pure nonequilibrium BCS state at all times.   
In fact, we find that the steady-state Mattis-Bardeen formulae for phase II 
are identical to the equilibrium finite-temperature results with 
$\Delta(T)$ replaced by $\Delta_\infty$ 
and 
the thermal quasiparticle distribution replaced by the nonequilibrium 
one $n_k$. 
The real part of the Mattis-Bardeen optical conductivity for phase II interaction quenches is 
presented in Fig.~\ref{Fig:MB_RePart_phaseII}. 
The imaginary part of the optical conductivity exhibits a $\Omega^{-1}$ power law decay. 
The prefactor of the imaginary part depends on $\Delta_{\infty}$.
Results are shown in Fig.~\ref{Fig:MB_ImPart_phaseII}.
These Mattis-Bardeen results for nonequlibrium superconductors in phase II are consistent with the 
experimental observations via optical measurements \cite{Matsunaga2013,Matsunaga2015}.

For phase I (strong deformation), the optical conductivity $\sigma_I(\Omega)$
 becomes indistinguishable from the 
normal metal. Namely, $\text{Re}[\sigma_{I}(\Omega)]=\sigma_N$ and $\text{Im}[\sigma_{I}(\Omega)]=0$. 
The approach to the phase II-I boundary from within phase II is qualitatively similar
to an equilibrium superconductor with $T \rightarrow T_c$, 
see Figs.~\ref{Fig:MB_RePart_phaseII} and \ref{Fig:MB_ImPart_phaseII}. 
Although $\Delta_\infty$ vanishes, it is known that phase I is not equivalent to a zero temperature
metallic state or any thermal ensemble; instead, it is a pure state with anomalous coherences, 
e.g., $\delta_k(t+t') \equiv \langle c_{-\vex{k} \downarrow}(t) \, c_{\vex{k} \uparrow}(t') \rangle \neq 0$. 
The order parameter $\Delta(t)$ vanishes due to dephasing between the different momenta contributing 
to Eq.~(\ref{Eq:GapEqn}). 
We also show that the superfluid density (defined via the London penetration depth) vanishes in phase I;
see Appendix~\ref{Sec:App:MPI}.
Our results for the Mattis-Bardeen conductivity and superfluid density
indicate that a different type of measurement 
(e.g., momentum-resolved) 
is
required to reveal the anomalous coherences that persist in phase I.


\section{Pump: Twisting Anderson pseudospins\label{Sec:pump}}

In the THz pump-probe experiments \cite{Matsunaga2013,Matsunaga2015}, a pump pulse 
is injected into a superconductor (with ground state BCS gap $\Delta_f \sim$ 0.5 THz) at the beginning. 
If we consider a system at zero temperature, then the superconducting state is 
deformed from the ground state by the pulse. 
After the light exposure, the new state evolves under the {\it original} BCS Hamiltonian, given by
Eq.~(\ref{Eq:BCS_MFT}), with $\vex{A}(t) = 0$.
In this section, the effect of the pump pulse is discussed in detail. 
The coupling between light to the superconducting order parameter can only arise 
in the nonlinear order \cite{Varma2002,Pekker2015}. 
We also discuss the connection between $\Delta_\infty$ and the twisting of the Anderson
pseudospin texture by the pump pulse.

\subsection{Light-superconductor coupling}

The electric field of the pump pulse 
in the self-consistent Hamiltonian given by Eq.~(\ref{Eq:BCS_MFT})
is given by 
$\vex{E}(t) = - (1/c) \, d \vex{A}(t) / d t$. 
We choose a Gaussian envelope, 
\begin{align}\label{PumpDetail}
	\vex{A}(t)
	=
	\hat{n}
	\,
	A
	\,
	\exp\left[-\frac{8\Delta_f^2(t-\pi/\Delta_f)^2}{\pi^2}\right] 
\end{align}
for $0\le t\le 2\pi/\Delta_f$, and zero outside of this window. 
Here, $\hat{n}$ is the unit polarization vector and $A$ is the 
peak
amplitude of the vector potential. 
The detailed shape of the pulse only modifies the results quantitatively. 
We select a particular pulse profile to simulate the experimental setup in Ref.~\onlinecite{Matsunaga2013}. 
Different from the previous numerical Bogoliubov approaches \cite{PapenKort2007,PapenKort2008,Krull2014}, 
we assume that the linearly polarized THz pump pulse excitation acts on the system uniformly due to the 
large wavelength of THz light ($\approx 300$ $\mu$m). 
If the initial state is a parity-symmetric BCS ground state, then we 
can rewrite 
the Anderson pseudospin Hamiltonian in Eq.~(\ref{Eq:MF_H}) using a different $\vec{B}$ field \cite{Tsuji2015},
\begin{align}\label{Eq:B_field_Light_SC}
	\vec{B}_{\text{pump},\vex{k}}
	=
	-
	2\left(
		\frac{
			\tilde\varepsilon_{\vex{k}-\frac{e}{c}\vex{A}}
			+
			\tilde\varepsilon_{\vex{k}+\frac{e}{c}\vex{A}}
		}{2}
		\hat{z}
		+
		\Delta_x \, \hat{x} + \Delta_y \, \hat{y}
	\right)\!,
\end{align}
where we have expressed $\vec{B}$ in terms of a symmetric combination that only admits even powers of $\vex{A}$. 
The linear coupling vanishes, as expected for a clean single-band superconductor \cite{Varma2002,Pekker2015}. 
For a 2D dispersion with 
$\varepsilon_{\vex{k}} = \varepsilon_{-\vex{k}}$, 
we expand 
$(
	\tilde\varepsilon_{\vex{k}-\frac{e}{c}\vex{A}}
	+
	\tilde\varepsilon_{\vex{k}+\frac{e}{c}\vex{A}}
)/2$ 
to the $\vex{A}^2$ order,
\begin{align}\label{Eq:dispersion_and_A}
	\frac{
		\tilde\varepsilon_{\vex{k}-\frac{e}{c}\vex{A}}
		+
		\tilde\varepsilon_{\vex{k}+\frac{e}{c}\vex{A}}}{2}
		\approx
		\tilde{\varepsilon}_{\vex{k}}
		+
		\frac{e^2}{2 c^2}\sum_{a,b=x,y}A_aA_b\partial_{k_a}\partial_{k_b}\tilde{\varepsilon}_{\vex{k}}.
\end{align}
For a pure isotropic quadratic dispersion, 
the effect of the light can be viewed as a time-dependent chemical potential shift 
that
does not alter the amplitude of the order parameter. 
It is therefore crucial to adopt a dispersion with energy-dependent curvature. 
Throughout this paper we study a square lattice nearest-neighbor tight-binding model 
at quarter filling \cite{quarter_filling}. The dispersion relation 
is 
$
	\varepsilon^{(\text{SL})}_{\vex{k}}
	=
	-2J
	\left[
		\cos\left(k_x a\right) + \cos\left(k_y a\right)
	\right]
$ 
where $J>0$ is the nearest-neighbor hopping strength and $a$ is the lattice constant. 
Equation~(\ref{Eq:dispersion_and_A}) becomes
\begin{align}\label{Eq:dispersion_and_A_SL}
	&\frac{
		\tilde\varepsilon^{(\text{SL})}_{\vex{k}-\frac{e}{c}\vex{A}}
		+
		\tilde\varepsilon^{(\text{SL})}_{\vex{k}+\frac{e}{c}\vex{A}}
	}{2}
	\approx
	\varepsilon^{(\text{SL})}_{\vex{k}}
	+
	\tilde{A}^2\mathcal{F}_{\vex{k}}(\alpha),
\\
	\label{Eq:F_k}
	&\mathcal{F}_{\vex{k}}(\alpha)
	=
	\left[
		\cos^2\left(\alpha\right)\cos\left(k_xa\right)
		+
		\sin^2\left(\alpha\right)\cos\left(k_ya\right)
	\right],
\end{align}
where $\alpha$ is the relative angle between the polarization axis of $\vex{A}$ and the $x$ direction, and 
$\tilde{A}^2$ defined by Eq.~(\ref{tilA2Def})
is the pump energy that characterizes the strength of light-superconductor coupling. 

The function $\mathcal{F}_{\vex{k}}(\alpha)$ encodes the anisotropy of the dispersion 
curvature, which depends on the lattice structure. 
The effect of such coupling induces a momentum-dependent $z$-component {\it magnetic} field for the Anderson pseudospins. 
The Anderson spins with different momenta precess differently under the light exposure. 
In all cases, the time-evolving order parameter reaches a steady value $\Delta(t) \rightarrow \Delta_\infty$
and  
$\Delta_{\infty}<\Delta_f$, 
where $\Delta_f$ is the ground state BCS gap. 
The results are insensitive to $\Delta_f/J$ (See Fig.~\ref{Fig:Phase_Diagram}). 
This suggests that the quadratic coupling to $\vex{A}$ captures the mechanism of the pump-pulse quench.
The values of $\Delta_{\infty}$ depend strongly on the polarization angle. 
In Fig.~\ref{Fig:Order_parameter_Lax1000}, polarization angles $\alpha=0$ and $\alpha=\pi/8$ give 
quantitatively different but qualitatively similar results. 
The angular dependence is not generic but is determined by the underlying lattice structure. 
We do not expect the same angular dependence in the THz pump-probe experiments \cite{Matsunaga2013,Matsunaga2015}.

\subsection{Twisted spin configurations and $\Delta_{\infty}$}

After the pump pulse, the real time dynamics of the superconducting order parameters in Fig.~\ref{Fig:Order_parameter_L2000_alpha0} 
exhibit behavior consistent with interaction quenches 
\cite{Barankov2004Collective_Rabi,Yuzbashyan2005PRB,Yuzbashyan2005JPA,Barankov2006,Yuzbashyan2006Relaxation,Yuzbashyan2006Dynamical_Vanishing}
in phases II ($\Delta_\infty \neq 0$) and phase I ($\Delta_\infty \rightarrow 0$).  
In Appendix~\ref{Sec:App:Lax} we construct the Lax vector and spectral polynomial 
that can be employed to determine the Lax roots of the post--pulse-quench Anderson pseudospin texture. 
We find two possible situations, zero (phase I) or one pairs (phase II) of isolated roots. 
For the phase II, the values of the isolated roots 
$u_0^{\pm}=\mu_{\infty}\pm i\Delta_{\infty}$, 
where $\mu_{\infty}$ and $\Delta_{\infty}$ correspond to the asymptotic values of 
the nonequilibrium chemical potential and amplitude of the order parameter in the long time limit. 

In Fig.~\ref{Fig:Spin_and_Lax}, the Anderson pseudospin textures with different pump energies are plotted for 
a small $L=24$ square lattice system at quarter-filling. The main effect of the pump pulse is to twist
the Anderson pseudospins near the Fermi surface in the $xy$ plane. 
We can mimic this twisting with a simple construction that leads to 
$\Delta_{\infty}<\Delta_f$, discussed in Appendix.~\ref{Sec:App:Lax}.
We have also confirmed that the isolated root predictions with the numerical simulations 
of the real-time pseudospin dynamics. Due to finite size effects, it is difficult to locate the 
phase boundary between phase II and phase I from the isolated roots alone. 
We numerically simulate systems with linear sizes $L=1000$, $2000$, and $4000$. 
The pump-pulse quench phase diagram is given in Fig.~\ref{Fig:Phase_Diagram}.

Some numerical evidence for vanishing $\Delta(t)$ in THz pumped 
self-consistent mean field dynamics was previously reported in Ref.~\cite{Papenkort2009},
although this was not interpreted in terms of phase I. 
Moreover, the nature of the excitation in that work was different, since
the ultrashort pump pulse assumed in \cite{Papenkort2009} had center frequency $\Omega$ 
greater than the ground state optical gap, $\Omega > 2 \Delta_f$.
Our pump pulse in Eq.~(\ref{PumpDetail}) has center frequency roughly equal to $\Delta_f$, 
and most of the spectral weight falls below $2 \Delta_f$ (similar to the experiment \cite{Matsunaga2013}). 
Our idea is that by exciting the ``Higgs dynamics'' entirely through the band nonlinearity, 
we are driving Anderson pseudospins in a way that would minimize linear absorption in a real, 
non--mean-field superconductor and maximize coherence. Understanding pair-breaking 
and decoherence by incorporating integrability-breaking terms would be necessary to prove this argument, which is
an important avenue for future work.
Our identification of the pump-quench induced phase II-I boundary with the disappearance
of the isolated Lax root pair unifies the interaction and pump-quenches within the 
Lax classification \cite{Yuzbashyan2006Relaxation} of BCS dynamics.


\section{Probe: Optical conductivity\label{Sec:Probe}}

In this section we extend the Mattis-Bardeen formulae \cite{Mattis1958} for the real 
and imaginary parts of the optical conductivity of a dirty superconductor
to the phase-II and phase-I nonequilibrium steady states.

\subsection{Brief review: Mattis-Bardeen formula}

We start with a brief review of the Mattis-Bardeen formula \cite{Mattis1958} for a superconductor at zero temperature. 
In the presence of elastic impurity scattering, momentum conservation is violated but energy 
conservation is preserved. 
Instead of deriving the explicit expression of the current-current correlation function 
in the presence of disorder, Mattis and Bardeen computed the ratio of the ac conductivity of a superconductor 
to the constant conductivity of a normal metal. The zero-temperature Mattis-Bardeen formula for the ac conductivity is
\begin{align}
	\nonumber
	\sigma_{\text{MB}}(\Omega)
	=&
	-
	\frac{i\sigma_N}{\pi\Omega}
	\!\int\! 
	d\varepsilon_1 d\varepsilon_2
	\left(u^*_1 u_1 v^*_2 v_2 - u^*_1 v_1 v^*_2 u_2\right)
	\\
\label{Eq:MB_Cond}
	&\times\!
	\left(	
		\frac{1}{\Omega\!+\!E_1\!+\!E_2\!+\!i\eta}
		\!
		-
		\!
		\frac{1}{\Omega\!-\!E_1\!-\!E_2\!+\!i\eta}
	\right)\!,\!\!
\end{align}
where $\sigma_N$ is the conductivity for the normal metal, $u_a$ and $v_a$ are the ground state coherence factors for 
single particle energy $\varepsilon_a$, 
$
	E_a
	\equiv
	\sqrt{\left(\varepsilon_a-\mu\right)^2+\Delta_f^2}
$, and $\eta$ is an infinitesimal positive number. 
Equation~(\ref{Eq:MB_Cond}) assumes a constant density of states. This generalization can be understood as follows. 
In the absence of the translational invariance, one can replace the labels $\vex{k}$ and $\vex{k}+\vex{q}$ in 
Eq.~(\ref{Eq:GS_pol}) by independent energy labels $\varepsilon_1$ and $\varepsilon_2$. 
The matrix elements of the current operators are randomized, and absorbed into the normal
state conductivity $\sigma_N$. This level of approximation neglects quantum interference
corrections. 

The real part of Eq.~(\ref{Eq:MB_Cond}) shows a gap for $\Omega<2\Delta_f$, 
an essential signature of zero temperature superconductivity \cite{Tinkham1996}. 
In order to evaluate the imaginary part of Eq.~(\ref{Eq:MB_Cond}), one has to 
regularize the formula by subtracting the normal metal contribution at zero frequency \cite{Mattis1958}. 
The result can be approximated by 
$
\sigma_{\text{MB}} / \sigma_N
\simeq
\pi\Delta_f/\Omega$ for $\Omega \ll 2\Delta_f$; 
$\lim_{\Omega \rightarrow 0} \left[-i \Omega \, \sigma_{\text{MB}}(\Omega)\right]$ is proportional to 
the superfluid density. 
Despite the simplifying assumptions made to obtain the Mattis-Bardeen formula, 
the finite temperature version 
is consistent with experimental observations for BCS superconductors \cite{Tinkham1996}.

\subsection{Weak quench: Phase II}

The long-time steady state of phase II can be expressed in terms of time-dependent coherence factors. 
These satisfy the asymptotic Bogoliubov-de Gennes equation,
\begin{align}
i\frac{d}{dt}
\left[
\begin{array}{c}
u_{\vex{k}}(t)\\
v_{\vex{k}}(t)
\end{array}
\right]
=\left[
\begin{array}{cc}
-\xi_{\vex{k}} & \Delta_{\infty}\\
\Delta_{\infty} & \xi_{\vex{k}}
\end{array}
\right]
\left[
\begin{array}{c}
u_{\vex{k}}(t)\\
v_{\vex{k}}(t)
\end{array}
\right],
\end{align}
where $\xi_{\vex{k}}=\varepsilon_{\vex{k}}-\mu_{\infty}$, 
and
$\mu_{\infty}$ and $\Delta_{\infty}$ 
are the long time asymptotic values of the
nonequilibrium
chemical potential and the order parameter, respectively.

The state is a pure BCS product wavefunction at all times, 
\begin{align}\label{Eq:BCS_T_wavefcn}
|\Psi_{II}(t)\rangle=\prod_{\vex{k}}\left[u_{\vex{k}}(t)+v_{\vex{k}}(t)s_{\vex{k}}^+\right]|0\rangle.
\end{align}
The time-dependent coherence factors are
\bsub\label{PhaseIICohFac}
\begin{align}
\label{Eq:u_TD}u_{\vex{k}}(t)&=\sqrt{1-n_{\vex{k}}}\,u_{\vex{k}}^{(0)}e^{i\mathcal{E}_{\vex{k}}t}-\sqrt{n_{\vex{k}}}\,v_{\vex{k}}^{(0)}e^{-i\mathcal{E}_{\vex{k}}t},\\
\label{Eq:v_TD}v_{\vex{k}}(t)&=\sqrt{1-n_{\vex{k}}}\,v_{\vex{k}}^{(0)}e^{i\mathcal{E}_{\vex{k}}t}+\sqrt{n_{\vex{k}}}\,u_{\vex{k}}^{(0)}e^{-i\mathcal{E}_{\vex{k}}t},
\end{align}
\esub
where $n_{\vex{k}}$ is the nonequilibrium quasiparticle distribution function (see Appendix~\ref{Sec:App:MBC})
and 
$\mathcal{E}_{\vex{k}}\equiv\sqrt{\xi_{\vex{k}}^2+\Delta_{\infty}^2}$. 
The time-independent coherence factors are
\begin{align}\label{Eq:u_v_TID}
u_{\vex{k}}^{(0)}\equiv \sqrt{\frac{\mathcal{E}_{\vex{k}}+\xi_{\vex{k}}}{2\mathcal{E}_{\vex{k}}}},\,\,\,\,v_{\vex{k}}^{(0)}\equiv -\sqrt{\frac{\mathcal{E}_{\vex{k}}-\xi_{\vex{k}}}{2\mathcal{E}_{\vex{k}}}}.
\end{align}

On can compute the two-time paramagnetic current polarization function 
for any BCS state of the form in Eq.~(\ref{Eq:BCS_T_wavefcn}); the result is given by 
Eq.~(\ref{Eq:TD_Polarization}). 
The current polarization function $\Pi(t,t';\vex{q})$ corresponding to Eq.~(\ref{Eq:BCS_T_wavefcn}) 
is a function separately of $t$ and $t'$. 
For the optical conductivity in the long-time asymptotic state, 
we only retain terms that can be expressed as functions of $t-t'$;
terms that oscillate with $(t + t')/2$ average out.  
The Mattis-Bardeen version of the ac conductivity then follows,
\bsub\label{MB-II}
\begin{align}
\sigma_{\text{II}}=&\sigma_{\text{II},a}+\sigma_{\text{II},b}\\[2mm]
\nonumber\sigma_{\text{II},a}=&-\frac{i\sigma_N}{\pi\Omega}\int d\varepsilon_1 d\varepsilon_2\\
\nonumber&\!\times\!\left\{\!\begin{array}{c}
\!\left(1-n_1\right)\left(1-n_2\right)\\
\times\left[u_1^{(0)}u_1^{(0)}
v_2^{(0)}v_2^{(0)}-u_1^{(0)}v_1^{(0)}
v_2^{(0)}u_2^{(0)}\right]\\[2mm]
-n_1n_2\left[v_1^{(0)}v_1^{(0)}u_2^{(0)}u_2^{(0)}-v_1^{(0)}u_1^{(0)}
\,u_2^{(0)}v_2^{(0)}\right]
\end{array}\!
\right\}\\
&\!\times\!\left[\!\frac{1}{\Omega+i\eta+\mathcal{E}_1+\mathcal{E}_2}-\frac{1}{\Omega+i\eta-\mathcal{E}_1-\mathcal{E}_2}\!\right],\\[2mm]
\nonumber\sigma_{\text{II},b}=&-\frac{i\sigma_N}{\pi\Omega}\int d\varepsilon_1 d\varepsilon_2
\left(1-n_1\right)n_2\\
\nonumber&\times\left[\begin{array}{c}
u_1^{(0)}u_1^{(0)}u_2^{(0)}u_2^{(0)}+v_1^{(0)}v_1^{(0)}v_2^{(0)}v_2^{(0)}\\[2mm]
+2v_1^{(0)}u_1^{(0)}
v_2^{(0)}u_2^{(0)}
\end{array}
\right]\\
&\times\left[\frac{1}{\Omega+i\eta+\mathcal{E}_1-\mathcal{E}_2}-\frac{1}{\Omega+i\eta-\mathcal{E}_1+\mathcal{E}_2}\right]\!.\!\!
\end{align}
\esub
Here
\begin{align}
\begin{gathered}
	\mathcal{E}_a \equiv \sqrt{\varepsilon_a^2 + \Delta_\infty^2}, 
	\\
	u_a \equiv \sqrt{\frac{\mathcal{E}_{a}+\varepsilon_a}{2\mathcal{E}_{a}}},
	\quad
	v_a \equiv -\sqrt{\frac{\mathcal{E}_{a} - \varepsilon_a}{2\mathcal{E}_{a}}},
\end{gathered}
\end{align}	
and $n_a = n(\varepsilon_a)$ is the nonequilibrium quasiparticle distribution function.
The energy ($\varepsilon_{1,2}$) integrations each run over the real line; the chemical potential can
be set equal to zero for the constant (disorder-averaged) density of states. 

The Mattis-Bardeen conductivity consists of two components. 
$\sigma_{\text{II},a}$ contains ``ground state Cooper pair'' and ``excited state Cooper pair'' 
contributions proportional to $(1 - n_1)(1 - n_2)$ and $n_1 n_2$, respectively.
The real part of $\sigma_{\text{II},a}$ is zero for $\Omega<2\Delta_{\infty}$, while  
$\sigma_{\text{II},b}$ contributes to all frequencies. 
In fact, Eq.~(\ref{MB-II}) turns out to be identical to the form obtained in thermal 
equilibrium at temperature $T$ \cite{Mattis1958} if we replace $\Delta(T) \rightarrow \Delta_\infty$
and the thermal quasiparticle distribution function by the nonequilibrium one $n_k$;
see Eqs.~(\ref{MB-II-Eq1})--(\ref{MB-II-Eq4}).

The real and imaginary parts of the ac conductivity for asymptotic phase II states are summarized in 
Figs.~\ref{Fig:MB_RePart_phaseII} and \ref{Fig:MB_ImPart_phaseII}. The detailed calculations are 
discussed in Appendix~\ref{Sec:App:MBC}. 
Since the results depend upon both $\Delta_\infty$ and the full nonequilibrium distribution function
$n_k$, we employ the exact analytical result
 for the latter that applies to instantaneous
interaction quenches. We expect that the results are qualitatively unchanged for the
THz pump quench. 
For interaction strength quenches, it is useful to introduce the 
quench parameter $\beta$ to locate the position in the phase diagram. 
This is defined via 
\begin{align}\label{betaDef}
	\beta
	\equiv
	2\left(\frac{1}{G_f}-\frac{1}{G_i}\right),
\end{align}
where $G_i$ ($G_f$) is the interaction strength of the pre-quench (post-quench) Hamiltonian.
For a 2D particle-hole symmetric superconductor with a uniform density of states, $\beta_c=\pi$ 
is the critical point that separates phase I ($\beta >\beta_c$) and phase II \cite{Barankov2006,Yuzbashyan2006Dynamical_Vanishing}.
The real part of the ac conductivity is very similar to that of a finite temperature superconductor with optical gap $2\Delta(T)$
replaced by $2\Delta_{\infty}$. 
With a quench parameter $\beta<\beta_c$, a peak arises at zero frequency, identical to the equilibrium situation 
with $T < T_c$.
When the quench strength $\beta$ approaches $\beta_c$, the ac conductivity converges to the normal metal result. 
The imaginary part of the conductivity exhibits $~1/\Omega$ behavior for $\Omega \ll 2\Delta_{\infty}$. 
We fit the imaginary part to
\begin{align}
	\text{Im}[\sigma_{\text{II}}(\Omega)]/\sigma_N
	=
	\frac{C(\beta) \, \Delta_{\infty}}{\Omega}.
\end{align} 
In Fig.~\ref{Fig:MB_ImPart_phaseII}, the prefactor $C(\beta)$ as a function of the quench parameter $\beta$ is 
plotted. For $0<\beta\ll \beta_c$, $C(\beta)\approx\pi$ (which is the equilibrium result for $T \ll T_c$). 
$C(\beta)$ deviates from $\pi$ as $\beta$ approaches $\beta_c$; the exact behavior depends upon the explicit
form of the nonequilibrium quasiparticle distribution function $n(\varepsilon)$.

\subsection{Strong quench: Phase I \label{Sec: PhaseIOC}}

For strong quenches, the order parameter vanishes to zero dynamically. 
In Eq.~(\ref{Eq:u_v_TID}), $\mathcal{E}_{\vex{k}}$ reduces to $|\xi_{\vex{k}}|$. Moreover, 
the time-dependent coherence factors in Eqs.~(\ref{Eq:u_TD}) and (\ref{Eq:v_TD}) become
\bsub\label{Eq:u_v_PhaseI}
\begin{align}
u_{\vex{k}}(t)&=
\sqrt{1-n_{\vex{k}}}\,e^{i\xi_{\vex{k}}t},
\\
v_{\vex{k}}(t)&=
\sqrt{n_{\vex{k}}}\,e^{-i\xi_{\vex{k}}t}.
\end{align}
\esub

The Mattis-Bardeen conductivity for phase I is
\begin{align}\label{MB-I}
	\nonumber
	\sigma_{\text{I}}(\Omega)=&-\frac{i\sigma_N}{\pi\Omega}\int d\varepsilon_1 d\varepsilon_2
	\, (1 - n_1) \, n_2 \\
	&\times\left[\frac{1}{\Omega+i\eta+\varepsilon_1-\varepsilon_2}-\frac{1}{\Omega+i\eta-\varepsilon_1+\varepsilon_2}\right]\!,
	\!\!
\end{align}
where we have used the same short-hand notations as in Eq.~(\ref{MB-II}).

Equation~(\ref{MB-I}) depends only on the nonequilibrium distribution function $n(\varepsilon_a)$. 
For an interaction quench, this is known explicitly and takes the same form as in phase II \cite{Yuzbashyan2006Dynamical_Vanishing}. 
One can show that the real part of Eq.~(\ref{MB-I}) is equal to $\sigma_N$ and
the imaginary part vanishes, so long as $n(\varepsilon)$ decays faster than $1/|\varepsilon|$
in the limit $|\varepsilon| \rightarrow \infty$. This is true for all phase I interaction quenches,
including the quench to zero pairing strength. 
In the latter case the result can be obtained analytically. 
Since the pump-pulse quench induces phase I mainly by twisting pseudospins near
the Fermi surface, and does so at much lower energy densities than an interaction
quench (demonstrated in Sec.~\ref{Sec:Thermalization}, below), we unfortunately conclude that the optical conductivity 
will not distinguish a thermalized normal metallic state from the quantum coherent
phase I of gapless superconductivity.

\subsection{Relation to experiments}

In the pump-probe experiments \cite{Matsunaga2013,Matsunaga2015}, the value of the superconducting 
order parameter $\Delta(t)$ was extracted from the kink in the real part of the measured optical conductivity. 
Our results 
(Fig.~\ref{Fig:MB_RePart_phaseII})
confirmed that this interpretation is consistent for the coherent nonequilibrium
steady states discussed here. 
We only compute the long time asymptotic behavior in this work, but the full two-time 
paramagnetic current polarization function necessary to monitor linear response at all times
can be determined using the Mattis-Bardeen version of Eq.~(\ref{Eq:TD_Polarization}). 
We have neglected inelastic processes such as electron--acoustic-phonon scattering 
that are ultimately responsible for thermalization.


\section{Internal energy of nonequilibrium states\label{Sec:Thermalization}}

\begin{figure}[b]
\centering
\includegraphics[width=0.4\textwidth]{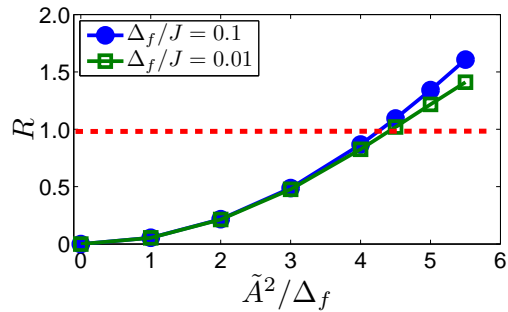}
\caption{The $R$-ratio for the pump-pulse quench. 
We consider the square lattice with linear system size $L=1000$ at quarter filling. 
The red dashed line marks $R = 1$. The phase II-I boundary is at $\tilde{A}^2/\Delta_f\approx 5.5$ 
($\tilde{A}^2/\Delta_f\approx 4.5$) for $\Delta_f/J=0.1$ ($\Delta_f/J=0.01$). 
In particular, the ratio $R\approx 1$ at the phase boundary 
($\tilde{A}^2/\Delta_f\approx 4.5$) for $\Delta_f/J=0.01$. 
All the phase I states show $R>1$. 
The polarization angle $\alpha=0$.
}
\label{Fig:pump_R}
\end{figure}

\begin{figure}[b]
\centering
\includegraphics[width=0.4\textwidth]{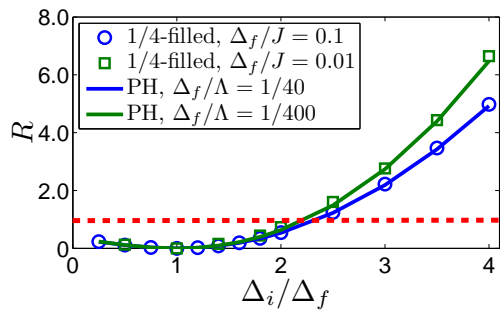}
\caption{The $R$-ratio for the interaction quench. 
The blue circles (green squares) correspond to a quarter-filled square lattice with $\Delta_f/J=0.1$ 
($\Delta_f/J=0.01$). 
$J$ is the strength of the nearest-neighbor hopping. The blue line (green line) corresponds to a 2D 
particle-hole symmetric model with $\Delta_f/\Lambda=1/40$ ($\Delta_f/\Lambda=1/400$).
In the latter, the single particle energies are uniformly distributed over $-\Lambda$ to $\Lambda$. 
The red dashed line marks $R=1$.
The phase II-I boundary is at 
$\Delta_i / \Delta_f \approx 4.6$ 
($\Delta_i / \Delta_f=e^{\pi/2}\approx 4.8$) 
for the quarter-filled square lattice (2D particle-hole symmetric model). 
All the data plotted here belong to phase II, and there is a sizable range with $R > 1$. 
All the phase I states show $R > 1$ in the interaction quench.
	}
\label{Fig:Int_qu_R}
\end{figure}

So far, we have only considered the post-pump dynamics of an 
integrable model for nonequilibrium superconductivity, i.e., 
time evolution according to the reduced BCS Hamiltonian in Eq.~(\ref{Eq:And_spin_H}).
Real superconductors
contain many integrability-breaking perturbations such as inelastic 
electron--optical-phonon 
and 
electron--acoustic-phonon 
scattering, as well as 
Cooper pair breaking due to residual quasiparticle interactions.
Thermalization to an equilibrium state always occurs on sufficiently long time scales; 
the novelty of the ultrafast ``pump quench'' in \cite{Matsunaga2013} is the ability to 
probe many-body dynamics on time scales shorter than this (i.e., 
within the pre-thermalization plateau). 
Of recent interest \cite{Basko2006,Nandkishore2014,Schreiber2015} are two different 
thermalization schemes: 
(a) the system thermalizes as a generic (nonintegrable), but closed system,
and 
(b) the system equilibrates with the external heat bath. 

Here we pose the following interesting thermalization scenario.  
Suppose the pump-quench--induced nonequilibrium superconductor thermalizes 
to an equilibrium state \emph{before} the external bath acts to absorb
the excess heat. 
Since the pump dumps a large amount of energy into the system, the
thermalized state can be either an equilibrium superconductor with $T < T_c$,
or a normal metal. 
We can distinguish these alternatives 
by comparing the injected internal energy of the nonequilibrium superconducting pure state 
to the internal energy of the equilibrium superconductor at the critical temperature,
relative to the ground state (zero temperature) energy. 
We define the ``$R$-ratio'' as follows,
\begin{align}\label{Eq:Ratio_R}
	R
	\equiv
	\frac{
		\left\langle\hat{H}_f\right\rangle_{\Psi}
		-
		\left\langle\hat{H}_f\right\rangle_{0}
	}{
		\left\langle\hat{H}_f\right\rangle_{T_c}
		-
		\left\langle\hat{H}_f\right\rangle_{0}},
\end{align} 
where $\left\langle\hat{H}_f\right\rangle_{\Psi}$ denotes the expectation of 
the post-quench Hamiltonian in the pumped state, 
$\left\langle\hat{H}_f\right\rangle_{0}$ is the ground state energy of the post-quench Hamiltonian, 
and 
$\left\langle\hat{H}_f\right\rangle_{T_c}$ is the thermal average  
of the post-quench Hamiltonian 
(internal energy density)
at the critical temperature. 
The post-quench Hamiltonian $\hat{H}_f$ corresponds to 
Eq.~(\ref{Eq:And_spin_H}) with $\vex{A} = 0$. 
If the nonequilibrium state carries more energy than the energy of a critical superconductor 
($R > 1$), 
the system will thermalize to a normal state.
Otherwise, it will thermalize to a finite temperature superconductor ($R < 0$).

For the pump pulse quench, we plot the $R$-ratio as a function of the pump energy in 
Fig.~\ref{Fig:pump_R}. We found that all phase I states show $R > 1$. 
This indicates that thermalization in a closed system will not produce a superconducting 
equilibrium state. Some of the phase II states close to the phase boundary also show $R > 1$. 
The $R > 1$ region decreases for smaller $\Delta_f/J$. 
Interestingly, the phase II-I boundary shows $R\approx 1$ for the smallest studied $\Delta_f/J=0.01$. 

In order to better understand the internal energy dependence of the nonequilibrium phases, 
we also compute the $R$-ratio for two types of interaction quenches:
(a) 2D square lattice with quarter filling and 
(b) 2D uniform particle-hole symmetric model with a finite bandwidth $2\Lambda$. 
Here, 
$\left\langle\hat{H}_f\right\rangle_{\Psi}$ 
denotes the expectation value of the post-quench Hamiltonian in the pre-quench ground state.
The results are shown in Fig.~\ref{Fig:Int_qu_R}.
Although interaction quenches (a) and (b) give nearly identical values of $R$ throughout most of 
phase II, both exceed $R = 1$ far from the phase II-phase I boundary. 
In other words, the injected internal energy of interaction quenches approaching
phase I is \emph{much larger} than for the pump pulse quench. 
It means that the internal energy alone is not a good indicator for the dynamical boundaries between
the nonequilibrium phases. This is not surprising, because the integrable nonequilibrium BCS 
dynamics are constrained by an infinite number of conserved quantities. 

The pump-pulse quench reaches phase I at much lower energy densities than the interaction quench.
This is because the former strongly scrambles the Anderson pseudospin texture for 
states near the Fermi energy, but barely modifies those away from it; see Fig.~\ref{Fig:Spin_and_Lax}. 
On the other hand, the interaction quenches affect all pseudospins regardless of the 
single particle energy. The interaction quench injects more energy than minimally required to 
generate nontrivial nonequilibrium dynamics. 
Due to the lower internal energy, one should expect that the window for observing nonequilibrium 
dynamics is larger in the pump pulse versus interaction quench protocols. 
In addition, the minimal required $R$-ratio might encode information about the phase II-I boundary. 
It is possible that $R=1$ at the phase II-I boundary for the optimized initial state(s) 
without any ``redundant'' internal energy.


\section{Discussion and conclusion\label{Sec:Discussion}}

In this paper, 
inspired by the experiments in Refs.~\cite{Matsunaga2013,Matsunaga2015}
we have studied quantum quench dynamics following the 
exposure of a BCS superconductor to an intense, approximately monocycle 
THz pulse. 
We connected the ``pump-pulse quench'' protocol to the interaction quench protocol 
using the Lax reduction method \cite{Yuzbashyan2005PRB,Yuzbashyan2005JPA}. 
We constructed the quench phase diagram in Fig.~\ref{Fig:Phase_Diagram}. 
Ultrafast spectroscopy with intense THz sources opens a new vista for 
solid state physics: the study of coherent, many-body quantum dynamics. 

Due to the large wavelength ($\approx 300$ $\mu$m) of the THz light, 
we assumed that the pump-pulse light couples to the entire system uniformly in space. 
Such coupling can be described in the Anderson pseudospin language. 
We focused on the second order (in vector potential amplitude) coupling, which is a curvature-dependent 
$z$-component effective magnetic field. 
We simulated the pump-pulse quench on a square lattice model at quarter-filling. 
We showed the possibility of accessing a gapless phase of fluctuating superconductivity (phase I) 
by increasing the pump energy. 

To connect the pump-pulse quench to the existing experiments \cite{Matsunaga2012,Matsunaga2013,Matsunaga2015}, 
we estimate the parameters including bandwidth, superconducting gap, and pump energy. 
The bandwidth of NbN is roughly 9.53eV \cite{Mattheiss1972}. 
One can estimate the tight binding parameter $J \sim 1.2$ eV. 
The optical superconducting gap in NbN is $2\Delta=5.2$ meV. 
This gives $\Delta_f/J\sim 2 \times 10^{-3}$. 
In Ref.~\onlinecite{Matsunaga2013}, the reported pump energies vary from 
$\tilde{A}^2=1\Delta_{f}$--$5\Delta_f$ for generating nonequilibrium superconductivity 
in phase II. The phase diagram in Fig.~\ref{Fig:Phase_Diagram} shows weak dependence upon the
value of $\Delta_f/J$. Our predicted phase boundary for the quarter-filled 
2D square lattice model is $\tilde{A}^2\approx 4.5\Delta_f$ with the polarization angle $\alpha=0$. 
We expect that phase I can be generated with the comparable magnitude of pump energies in the THz pump-probe 
experiments.

Optical conductivity is the quantity measured in Refs.~\cite{Matsunaga2012,Matsunaga2013,Matsunaga2015}. 
We computed the Mattis-Bardeen optical conductivity for dirty superconductors in phases II and I. 
The phase II result is the same as the prediction for an equilibrium superconductor with $T < T_c$,
except that it depends on the asymptotic nonequilibrium order parameter $\Delta_\infty$ and the
nonequilibrium quasiparticle distribution function. These results
back up the gap-extraction procedures employed in the existing experiments \cite{Matsunaga2013,Matsunaga2015}. 
The phase I optical conductivity cannot be distinguished from a normal metal. 
In Appendix~\ref{Sec:App:MPI}, we show that the superfluid density (defined via the London penetration depth) 
also vanishes in phase I. 

We have assumed integrable post-pump dynamics throughout this work. 
Integrability-breaking processes induce thermalization and limit the time window 
(pre-thermalization plateau) for observing quench dynamics. 
Moreover, phonons \cite{Matsunaga2013,Krull2014,Kemper2015,Murakami2016,Sentef2016} and other coexisting collective modes \cite{Akbari2013,Moor2014,Fu2014,Cea2014,Dzero2015,Krull2016,Sentef2016arXiv} might modify the evolution in a superconductor. 
A detailed study of thermalization and decoherence out of the 
far-from-equilibrium superconductor states discussed here remains a challenging project for future work. 
Based on the results in Sec.~\ref{Sec:Thermalization}, the pump-pulse quench is a much more
efficient protocol for accessing nonlinear nonequilibrium dynamics, compared to interaction quenches. 
It would be interesting to try to reach other dynamical phases (e.g., the
quench-generated Floquet phase III) by engineering 
different ultrafast excitation schemes. One possibility would be to design a sequence 
of pump pulses of variable duration and delays. 
Realizing the pump-pulse quench protocol in an ultracold fermion condensate 
might also yield more understanding for THz pump-probe experiments.

\section*{Acknowledgements}

We thank Maxim Dzero, Junichiro Kono, Ryo Shimano, and Emil Yuzbashyan for helpful discussions.
This research was supported by the Welch Foundation grant No.~C-1809 and by 
NSF CAREER grant no.~DMR-1552327 (Y.-Z.C., Y.L., and M.S.F.). 
This work was also supported by NSF grant no.~DMR-1001240, 
through the Simons Investigator award of the Simons Foundation, 
and by a Center for Theoretical Quantum Matter postdoctoral fellowship at University of Colorado Boulder (Y.-Z.C.).

\appendix

\section{Lax construction and spectral polynomial; twisted pseudospins \label{Sec:App:Lax}}

For a classical system consisted of $N$ pseudospins, we define the Lax vector components \cite{Yuzbashyan2005PRB},
\begin{align}
L^{z}(u)&\equiv\sum_{j=1}^{N}\frac{s_j^z}{u-\varepsilon_j}-\frac{1}{G},\\
L^{x}(u)&\equiv\sum_{j=1}^{N}\frac{s_j^x}{u-\varepsilon_j},\,\,\,L^{y}(u)\equiv\sum_{j=1}^{N}\frac{s_j^y}{u-\varepsilon_j},
\end{align}
where $G$ is the strength of the BCS interaction in Eq.~(\ref{Eq:And_spin_H}) and $u$ denotes a complex parameter.
The Lax norm is defined as
\begin{align}
\vec{L}^2(u)=\left[L^x(u)\right]^2+\left[L^y(u)\right]^2+\left[L^z(u)\right]^2.
\end{align}
We impose the canonical Poisson bracket relation for the pseudospins,
\begin{align}
\left\{s^{a}_j,s^b_k\right\}=\delta_{j,k}\epsilon^{abc}s_j^c.
\end{align}
Based on the above condition, one can show that the Lax vector components satisfy
\begin{align}
\left\{L^a(u),L^b(v)\right\}&=-\epsilon^{abc}\left[\frac{L^c(u)-L^c(v)}{u-v}\right],\\
\left\{\vec{L}^2(u),\vec{L}^2(v)\right\}&=0.
\end{align}

We define the spectral polynomial
\begin{align}
\mathcal{Q}_{2N}(u)\equiv G^2\prod_{j}^{N}\left(u-\varepsilon_j\right)^2\vec{L}^2(u).
\end{align}
$\mathcal{Q}_{2N}(u)$ is a polynomial of degree $2N$ in the parameter $u$;
it is conserved under evolution by the BCS Hamiltonian in Eq.~(\ref{Eq:And_spin_H}) with $\vex{A} = 0$.
The spectral polynomial for the BCS ground state can be written as
\begin{align}
\mathcal{Q}_{2N}(u)&=\left(\frac{G}{2}\right)^2\left[\left(u-\mu\right)^2+|\Delta_f|^2\right]\mathcal{P}^2_{N-1}(u),\\
\mathcal{P}_{N-1}(u)&=\left[\prod_{j}\left(u-\varepsilon_j\right)\right]\mathcal{F}(u),		\label{PN-1}\\
\mathcal{F}(u)&=\sum_j\frac{1}{\left(u-\varepsilon_j\right)\sqrt{\tilde{\varepsilon}_j^2+|\Delta_f|^2}}.
\end{align}
There are $2(N-1)$ real roots and 2 complex roots in $\mathcal{Q}_{2N}$. The two complex roots (isolated roots) encode 
the ground state BCS gap $\Delta_f$,
\begin{align}
u^{\pm}_0=\mu\pm i|\Delta_f|.
\end{align}

For interaction quenches, it has been shown that most of the roots of the
spectral polynomial lie along the real axis in the thermodynamic limit. 
The complex roots (isolated roots)
determine the asymptotic long time behavior $\Delta(t)$ following the quench. 
The procedure used to extract this behavior is called Lax reduction \cite{Yuzbashyan2006Relaxation}. 
Interaction quenches from a BCS initial state exhibit 
0, 1, or 2 pairs of isolated roots, associated to exactly solvable   
0, 1, or 2 collective-pseudospin problems \cite{Yuzbashyan2006Relaxation,Barankov2006,Yuzbashyan2006Dynamical_Vanishing}.
These respectively correspond to phases I, II, and III discussed in Sec.~\ref{Sec: model, intq review}.  
In phase II, $\Delta_\infty$ is encoded in the single isolated root pair
$u_0^{\pm}=\mu_{\infty}\pm i\Delta_{\infty}$, where $\mu_{\infty}$ is the effective nonequilibrium chemical potential.

As demonstrated by the Anderson pseudospin textures exhibited in 
Fig.~\ref{Fig:Spin_and_Lax}, the main effect of the THz pump pulse 
is to twist pseudospins along the Fermi surface
in the $xy$ pseudospin plane. This ``scrambling'' induces an asymptotic  $\Delta_\infty$ which
is strictly less than the ground state gap $\Delta_f$, as shown in the dynamical phase diagram 
Fig.~\ref{Fig:Phase_Diagram}.

We can mimic the pump-induced twisting with the following simple construction.   
We consider a situation such that the $x$- and $y$-components of the pseudospin texture are 
deformed relative to the BCS ground state configuration.  We write down the Lax vector as a sum over equal energy sectors,
\begin{align}
\vec{L}(u)&\equiv\sum_{j=1}^N\frac{\vec{s}_j}{u-\varepsilon_j}-\frac{\hat{z}}{G}\\
&\rightarrow\sum_{n}\sum_{k_n}\frac{\vec{s}_{k_n}}{u-e_n}-\frac{\hat{z}}{G}=\sum_{n}\frac{\vec{M}_{n}}{u-e_n}-\frac{\hat{z}}{G},
\end{align}
where $e_n$ indicates the $n^{\mathrm{th}}$ distinct energy level, 
$\vec{s}_{k_n}$ is the $k_n^\mathrm{th}$ spin with energy $e_n$, 
and 
$\vec{M}_n$ is the net polarization of the spins with energy $e_n$.

The Lax vector for the twisted configuration is defined via 
\begin{align}
	L_{\text{t}}^{x,y}(u)&=\sum_n\frac{\lambda_n M_{n}^{x,y}}{u-e_n}
	\\
	&=\sum_n\sum_{k_n}\frac{1}{u-e_n}\left[\frac{-1}{2}\frac{\lambda_n \Delta^{x,y}}{\sqrt{\left(e_n-\mu\right)^2+|\Delta_f|^2}}\right],
	\nonumber
	\\
	L_{\text{t}}^z(u)&=L^z(u),
\end{align}
where $\lambda_n <1$ characterizes the degree of twisting in the sector of single energy $e_n$. 
For simplicity we take $\lambda_n = \lambda$ (independent of $e_n$). Then the spectral polynomial for
the twisted texture can be written as
\begin{align}
\mathcal{Q}_{2N}(u)=\left(\frac{G}{2}\right)\left[\left(u-\mu\right)^2+\lambda^2|\Delta_f|^2\right]\mathcal{P}^2_{N-1}(u),
\end{align}
where $\mathcal{P}_{N-1}(u)$ is the same degree $N-1$ polynomial that appears in the ground state
[Eq.~(\ref{PN-1})]. The isolated roots give 
$u_0^{\pm}=\mu\pm i\lambda|\Delta_f|$. 
The asymptotic value of the order parameter is therefore 
$\lambda|\Delta_f| < |\Delta_f|$. 
This suggests that one can achieve phase I (effective 0 spin problem with $\Delta_\infty \rightarrow 0$) 
by increasing the degree of the twisting.


\section{Linear response theory and Kubo formula}

We obtain the general expression for the Kubo conductivity associated to a generic, time-evolving pure BCS product
state in this section. In the continuum limit, the paramagnetic current operator is expressed as
\begin{align}
	\vex{J}(t,\vex{q})=
	-
	\frac{e}{m}\sum_{\vex{k},\sigma}\left(\vex{k}+\frac{\vex{q}}{2}\right)c^{\dagger}_{\vex{k}\sigma}(t)c_{\vex{k}+\vex{q}\sigma}(t),
\end{align}
where $m$ is the effective mass of electron.
Linear response theory gives
\begin{align}
	\langle J^{\mu}(t,\vex{q})\rangle
	=
	-
	\frac{1}{c}
	\int_{-\infty}^{t} dt' 
	\,
	\Pi^{\mu\nu}_R(t,t';\vex{q})
	\,
	A^{\mu}(t',\vex{q}),
\end{align}
where 
\begin{align}
	\Pi^{\mu\nu}_R(t,t';\vex{q})
	\equiv
	-i \left\langle
	\left[J^{\mu}(t,\vex{q}),J^{\nu}(t',-\vex{q}')\right]
	\right\rangle_0 \, \theta(t - t')
\end{align}
is the retarded paramagnetic current polarization function. 
$A^{\mu} = (c E^{\mu} / i \Omega) \, \exp(- i \Omega t)$ is the vector potential for a 
monochromatic source with frequency $\Omega$.
The expectation $\langle \cdots \rangle_0$ is performed in terms of the 
initial BCS ground state, while the Heisenberg picture current operators 
incorporate the effects of the pump pulse as well as the subsequent free 
time-evolution under the BCS Hamiltonian. 

In the ground state, $\Pi^{\mu\nu}_R(t,t';\vex{q})\equiv\Pi^{\mu\nu}_R(t-t',\vex{q})$ 
is translational invariant in time.
The retarded polarization function in the frequency-momentum space is \cite{MahanBook}
\begin{align}
\nonumber
\Pi^{\mu\nu}_R\!(\Omega,\vex{q})\!=\!&-2\left(\frac{e}{m}\right)^2\sum_{\vex{k}}\left(k^{\mu}+\frac{q^{\mu}}{2}\right)\left(k^{\nu}+\frac{q^{\nu}}{2}\right)\\
\nonumber&\!\times\!\left[u^*_{\vex{k}+\vex{q}}u_{\vex{k}+\vex{q}}v^*_{\vex{k}}v_{\vex{k}}
-u^*_{\vex{k}+\vex{q}}v_{\vex{k}+\vex{q}}v^*_{\vex{k}}u_{\vex{k}}\right]\\
\label{Eq:GS_pol}&\!\times\!\left[\!\frac{1}{\Omega\!+\!i\eta+\!E_{\vex{k}}\!+\!E_{\vex{k}+\vex{q}}}\!-\!\frac{1}{\Omega\!+\!i\eta-\!E_{\vex{k}}\!-\!E_{\vex{k}+\vex{q}}}\!\right].
\end{align}
The optical conductivity is
\begin{align}
	\sigma(\Omega)
	=
	\frac{i}{\Omega}
	\left[\lim\limits_{\vex{q}\rightarrow 0}\Pi^{xx}_R(\Omega,\vex{q})
	+
	\frac{e^2 n}{m}
	\right],
\end{align}
where the second term is the diamagnetic contribution. 
Since $\Pi^{\mu\nu}_R\!(\Omega,\vex{q} \rightarrow 0) = 0$, the
real part of the optical conductivity is exactly zero for a clean 
(single band) superconductor. 
Low-temperature superconductors typically reside in the dirty limit
due to the presence of quenched disorder. 
Relaxing momentum conservation in Eq.~(\ref{Eq:GS_pol}) leads to the Mattis-Bardeen formulae \cite{Mattis1958}.

For the BCS wavefunction in Eq.~(\ref{Eq:BCS_T_wavefcn})
with \emph{generic} time-dependent coherence factors,
the polarization function can be evaluated. 
The result is 
\begin{align}\label{Eq:TD_Polarization}
	\Pi^{\mu\nu}_R(t,t';\vex{q})
	=&
	-2i\left(\frac{e}{m}\right)^2\sum_{\vex{k};\sigma,\sigma'}
	\left(k^{\mu}+\frac{q^{\mu}}{2}\right)\left(k^{\nu}+\frac{q^{\nu}}{2}\right)
\nonumber\\
&\times
	\left[
	\begin{array}{c}
	u^*_{\vex{k}+\vex{q}}(t)u_{\vex{k}+\vex{q}}(t')v^*_{\vex{k}}(t)v_{\vex{k}}(t')\\[2mm]
	-u^*_{\vex{k}+\vex{q}}(t)v_{\vex{k}+\vex{q}}(t')v^*_{\vex{k}}(t)u_{\vex{k}}(t')\\
	-(t\leftrightarrow t')
	\end{array}
\right]
\nonumber\\
&\times
\theta(t - t').	
\end{align}
Again this vanishes for $\vex{q}\rightarrow 0$. 
Since it applies to any state of BCS form, Eq.~(\ref{Eq:TD_Polarization}) determines the
paramagnetic current polarization function for all three phases (I, II, III) of 
quench-induced nonequilibrium superconductivity, reviewed in Sec.~\ref{Sec: model, intq review}. 
In the asymptotic steady state (quasi-steady state) of phases II or I (phase III),
we can average $\Pi^{\mu\nu}_R(t,t';\vex{q})$ over the center-of-time $(t + t')/2$ to obtain
an effective formula that depends only on $t - t'$ (similar to the ground state). 
Incorporating disorder ala Mattis and Bardeen, one obtains the frequency-dependent 
optical conductivity for these nonequilibrium phases.


\section{Mattis-Bardeen conductivity in phases II and I \label{Sec:App:MBC}}

We express the nonequilibrium quasiparticle distribution function $n_{\vex{k}}$ in terms 
of a function $\gamma(\vex{k})$ via $n_{\vex{k}}=\left[1+\gamma(\vex{k})\right]/2$. 
For an interaction quench with strength $\beta$ [Eq.~(\ref{betaDef})] and initial
BCS gap $\Delta_i$, the latter is given by \cite{Yuzbashyan2006Dynamical_Vanishing,Yuzbashyan2015} 
\begin{align}\label{Eq:dist_fcn}
\begin{aligned}
	\gamma(\varepsilon_{\vex{k}})
	=&\,
	s(\varepsilon_{\vex{k}})
	\left|\gamma(\varepsilon_{\vex{k}})\right|,
	\\
	\left|\gamma(\varepsilon_{\vex{k}})\right|
	=&\,
	\sqrt{
		1
		-
		\frac{
			\left[
				\mathcal{N}(\varepsilon_{\vex{k}})
				-
				\sqrt{
				\mathcal{N}^2(\varepsilon_{\vex{k}})
				-
				\left[\frac{2\pi \tilde{\nu}(\varepsilon_{\vex{k}})\Delta_i\beta}{E_i(\varepsilon_{\vex{k}})}\right]^2
				}
				\,\right]}{2\left[\pi \tilde{\nu}(\varepsilon_{\vex{k}})
			\right]^2}
	},
\end{aligned}
\end{align}
where 
\[
	\tilde{\nu}(\varepsilon) \equiv \frac{\nu(\varepsilon)}{\nu(\varepsilon_F)}
\]
is the dimensionless density of states ($\varepsilon_F$ is the Fermi energy)
and
$
	E_i(\varepsilon)
	=
	\sqrt{\left(\varepsilon-\mu_i\right)^2+\Delta_i^2}
$, 	
where $\mu_i \simeq \varepsilon_F$ is the prequench chemical potential. 
The function $\mathcal{N}(\varepsilon)$ is given by 
\begin{align}
	\mathcal{N}(\varepsilon)
	\!
	=
	\!
	\left[
		\left(\varepsilon-\mu_i\right)f_R(\varepsilon)\!+\!\beta\right]^2
		\!\!
		+
		\!
		\left[\Delta_i f_R(\varepsilon)\right]^2
		\!\!
		+
		\!
		\pi^2 \tilde{\nu}^2(\varepsilon),
\end{align}
where
\begin{align}
	f_R(u)
	=
	\mathcal{P}
	\int d{\varepsilon}
	\frac{\tilde{\nu}(\varepsilon)}{(u-\varepsilon)\sqrt{\left(\varepsilon-\mu_i\right)^2+\Delta_i^2}}.
\end{align}
Finally, the function $s(\varepsilon) = \pm 1$ in Eq.~(\ref{Eq:dist_fcn}) determines the sign of 
$\gamma(\varepsilon)$. Eq.~(\ref{Eq:dist_fcn}) applies to interaction quenches in all three dynamical
phases I, II, and III, but the sign $s(\varepsilon) = \pm 1$  must be carefully determined for any 
quench such that $\gamma$ is a smooth function of $\varepsilon$. This is relevant for phase I quenches,
since $\gamma(\varepsilon) \rightarrow +1$ [$\gamma(\varepsilon) \rightarrow -1$] for states far below
(above) the Fermi energy. 

The Mattis-Bardeen version of Eq.~(\ref{Eq:TD_Polarization}) 
for the phase II coherence factors [Eq.~(\ref{PhaseIICohFac})] was 
given above by Eq.~(\ref{MB-II}). Using  
$\sigma_{\text{II}}=\sigma_{\text{II},a}+\sigma_{\text{II},b}$,
Eq.~(\ref{MB-II}) can be rewritten as 
\begin{widetext}
\begin{align}
\text{Re}\left[\sigma_{\text{II},a}\right](\Omega)=&\frac{\sigma_{\text{N}}}{8\Omega}\int d\varepsilon_1 d\varepsilon_2
\left[\gamma_1+\gamma_2\right]
\left[1-\frac{\Delta_{\infty}^2}{\mathcal{E}_1\mathcal{E}_2}\right]
\left[\delta\left(\Omega+\mathcal{E}_1+\mathcal{E}_2\right)-\delta\left(\Omega-\mathcal{E}_1-\mathcal{E}_2\right)\right],
\label{MB-II-Eq1}
\\
\text{Im}\left[\sigma_{\text{II},a}\right](\Omega)=&\frac{\sigma_{\text{N}}}{8\pi\Omega}\int d\varepsilon_1 d\varepsilon_2
\left[\gamma_1+\gamma_2\right]
\left[1-\frac{\Delta_{\infty}^2}{\mathcal{E}_1\mathcal{E}_2}\right]
\left[\frac{\Omega+\mathcal{E}_1+\mathcal{E}_2}{\left(\Omega+\mathcal{E}_1+\mathcal{E}_2\right)^2+\eta^2}-
\frac{\Omega-\mathcal{E}_1-\mathcal{E}_2}{\left(\Omega-\mathcal{E}_1-\mathcal{E}_2\right)^2+\eta^2}\right],
\label{MB-II-Eq2}
\\
\text{Re}\left[\sigma_{\text{II},b}\right](\Omega)=&\frac{\sigma_{\text{N}}}{8\Omega}\int d\varepsilon_1 d\varepsilon_2
\left[\gamma_1-\gamma_2\right]
\left[1+\frac{\Delta_{\infty}^2}{\mathcal{E}_1\mathcal{E}_2}\right]
\left[\delta\left(\Omega+\mathcal{E}_1-\mathcal{E}_2\right)-\delta\left(\Omega-\mathcal{E}_1+\mathcal{E}_2\right)\right],
\label{MB-II-Eq3}
\\
\text{Im}\left[\sigma_{\text{II},b}\right](\Omega)=&\frac{\sigma_{\text{N}}}{8\pi\Omega}\int d\varepsilon_1 d\varepsilon_2
\left[\gamma_1-\gamma_2\right]
\left[1+\frac{\Delta_{\infty}^2}{\mathcal{E}_1\mathcal{E}_2}\right]
\left[\frac{\Omega+\mathcal{E}_1-\mathcal{E}_2}{\left(\Omega+\mathcal{E}_1-\mathcal{E}_2\right)^2+\eta^2}
-\frac{\Omega-\mathcal{E}_1+\mathcal{E}_2}{\left(\Omega-\mathcal{E}_1+\mathcal{E}_2\right)^2+\eta^2}\right],
\label{MB-II-Eq4}
\end{align}
\end{widetext}
where short-hand notations have been used. E.g., $\mathcal{E}_a\equiv\mathcal{E}(\varepsilon_a)$ and $\gamma_a\equiv \gamma(\varepsilon_a)$ with $a=1,2$.
Eqs.~(\ref{MB-II-Eq1})--(\ref{MB-II-Eq4}) are identical to the
finite temperature $T$ Mattis-Bardeen result \cite{Mattis1958} if we replace 
$\Delta(T) \rightarrow \Delta_\infty$ and the thermal distribution function by 
$n(\varepsilon) = [1 + \gamma(\varepsilon)]/2$. 
The imaginary part of the conductivity in Eqs.~(\ref{MB-II-Eq2}) and (\ref{MB-II-Eq4}) 
contains a UV-divergent contribution that can be regularized by subtracting the normal metal contribution at $\Omega=0$ \cite{Mattis1958}. 
In the dirty limit captured by the Mattis-Bardeen formulae, the phase II distribution function 
$\gamma(\varepsilon)$ is taken to be that of a particle-hole symmetric metal 
with a constant density of states. This
gives Eq.~(\ref{Eq:dist_fcn}) with 
$\tilde{\nu}(\varepsilon) = 1$ 
and 
$s(\varepsilon) = -1$. 

For a phase I interaction quench, $\gamma(\varepsilon)$ is still given by Eq.~(\ref{Eq:dist_fcn}), 
but now we must take $s(\varepsilon) = -\sgn(\varepsilon)$. The Mattis-Bardeen formula Eq.~(\ref{MB-I}) 
is 
\begin{widetext}
\begin{align}
\text{Re}\left[\sigma_{\text{I}}\right](\Omega)=&\frac{\sigma_{\text{N}}}{4\Omega}\int d\varepsilon_1 d\varepsilon_2
\left[\gamma(\varepsilon_1)-\gamma(\varepsilon_2)\right]
\left[\delta\left(\Omega+\varepsilon_1-\varepsilon_2\right)-\delta\left(\Omega-\varepsilon_1+\varepsilon_2\right)\right],\\
\text{Im}\left[\sigma_{\text{I}}\right](\Omega)=&\frac{\sigma_{\text{N}}}{4\pi\Omega}\int d\varepsilon_1 d\varepsilon_2
\left[\gamma(\varepsilon_1)-\gamma(\varepsilon_2)\right]
\left[\frac{\Omega+\varepsilon_1-\varepsilon_2}{\left(\Omega+\varepsilon_1-\varepsilon_2\right)^2+\eta^2}
-\frac{\Omega-\varepsilon_1+\varepsilon_2}{\left(\Omega-\varepsilon_1+\varepsilon_2\right)^2+\eta^2}\right].
\end{align}
\end{widetext}
The imaginary part of the conductivity again needs to be regularized by subtracting the normal metal contribution at $\Omega=0$. 
In the quench to zero pairing strength [free fermion limit, $G_f=0$ and $\beta\rightarrow\infty$ in Eq.~(\ref{betaDef})], 
the phase I distribution function reduces to
\begin{align}
\lim\limits_{\beta\rightarrow\infty}\left[\gamma(\varepsilon)\right]=-\frac{\varepsilon-\mu_i}{\sqrt{\left(\varepsilon-\mu_i\right)^2+\Delta_i^2}}.
\end{align}
The free fermion quench results can be evaluated analytically. 
For the generic situation, one must evaluate the optical conductivity numerically. 
As discussed in the text, the phase I conductivity is identical to the conductivity of a normal state.


\section{Vanishing superfluid density in phase I \label{Sec:App:MPI}}

In this appendix, we show that the superfluid density 
that determines the London penetration depth vanishes in phase I.
This is consistent with the observation that the Mattis-Bardeen
optical conductivity is that of a normal metal 
(Secs.~\ref{Sec: OptI,II} and \ref{Sec: PhaseIOC}),
although the optical conductivity and Meissner effect obtain
from different (and in general non-commuting) limits of the
paramagnetic current-current correlation function. 

To be precise, we consider the linear response to a static vector
potential $\vex{A}(\vex{q})$ in the asymptotic post-quench steady state. 
We will compute the screening current in the clean 2D square lattice
model employed in the pump-pulse quench, but our results are more general. 
The current is 
\begin{widetext}
\begin{align}\label{London}
	\left\langle J^\mu(\vex{q}) \right\rangle
	=&\,
	-
	\frac{1}{c}
	\Pi_R^{\mu \nu}(\Omega = 0,\vex{q})
	\,
	A^\nu(\vex{q}) 
	+
	\left\langle J_{2}^\mu(\vex{q}) \right\rangle,
\\
	\left\langle J_{2}^\mu(\vex{q} \rightarrow 0) \right\rangle
	=&\,
	-
	\frac{J e^2}{c}
	\sum_{\sigma,\lambda}
	\int \frac{d^2 k}{(2 \pi)^2}
	\int_{-\infty}^\infty 
	d t \,
	\langle 
	c^\dagger_{\vex{k},\sigma}(t)
	\,
	c_{\vex{k},\sigma}(t)
	\rangle 
	\,
	\cos(\vex{k}\cdot\vex{V}_\lambda)
	\,
	V_\lambda^\mu
	\,
	V_\lambda^\nu
	\,
	A^\nu(t).
\end{align}
\end{widetext}
Here $J$ is the hopping strength and $V_\lambda^\mu$ is a nearest-neighbor vector
for the 2D square lattice; $1 \leq \lambda \leq 4$ indicates the four nearest-neighbor sites. 
The band energy is 
$
	\varepsilon^{(\text{SL})}_{\vex{k}} = - J \sum_{\lambda} \cos\left(\vex{k} \cdot \vex{V}_\lambda\right).
$
In a generic time-dependent BCS state [Eq.~(\ref{Eq:BCS_T_wavefcn})], 
the diamagnetic current $J_2^\mu$ can be evaluated using
\begin{align}
	\sum_{\sigma}
	\langle 
	c^\dagger_{\vex{k},\sigma}(t)
	\,
	c_{\vex{k},\sigma}(t)
	\rangle 
	=
	2 |v_{\vex{k}}(t)|^2.
\end{align}
In phase I, Eq.~(\ref{Eq:u_v_PhaseI}) implies that this is independent of time
and gives $|v_{\vex{k}}(t)|^2 = n_{\vex{k}}$. Here $n_{\vex{k}}$ is the average electron occupation 
of the states $\ket{\vex{k},\uparrow}$ and $\ket{-\vex{k},\downarrow}$. 
Thus the diamagnetic current is 
\begin{align}
	\left\langle J_{2}^\mu \right\rangle
	=&\,
	-
	\frac{2 e^2 J}{c}
	\sum_{\lambda}
	\int \frac{d^2 k}{(2 \pi)^2}
	n_{\vex{k}}
	\,
	\cos(\vex{k}\cdot\vex{V}_\lambda)
	\,
	V_\lambda^\mu
	\,
	V_\lambda^\nu
	\,
	A^\nu.
\end{align}

For a generic time-evolving BCS state, the 
retarded paramagnetic current-current correlation function
$\Pi_R^{\mu \nu}(t,t';\vex{q})$ is the lattice version
of Eq.~(\ref{Eq:TD_Polarization}). 
Using the phase I coherence factors [Eq.~(\ref{Eq:u_v_PhaseI})]
gives a function only of $(t - t')$. The zero frequency
Fourier transform on the square lattice is 
\begin{widetext}
\begin{align}
	\Pi_R^{\mu \nu}(\Omega = 0,\vex{q})
	=
	-
	2(e J)^2
	\sum_{\lambda,\lambda'}
	\int \frac{d^2 k}{(2 \pi)^2}
	V^\mu_\lambda 
	\,
	V^\nu_{\lambda'}
	\,
	\sin\left[\left(\vex{k} + \frac{\vex{q}}{2}\right)\cdot\vex{V}_\lambda\right]
	\,
	\sin\left[\left(\vex{k} + \frac{\vex{q}}{2}\right)\cdot\vex{V}_{\lambda'}\right]
	\,
	\left(
	n_{\vex{k}}
	-
	n_{\vex{k}+\vex{q}}
	\right)
	\,
	\frac{1}{\varepsilon^{(\text{SL})}_{\vex{k}+\vex{q}} - \varepsilon^{(\text{SL})}_{\vex{k}} }.
\end{align}

Now, we assume that $n_{\vex{k}} = n(\varepsilon^{(\text{SL})}_{\vex{k}})$ only.
This is certainly true for an interaction quench; we also 
expect it to hold for the pump-pulse quench, since the main 
effect of the pump is to twist Anderson pseudospins in the 
$xy$ plane. 
In the $\vex{q} \rightarrow 0$ limit, this gives
\begin{align}
	\Pi_R^{\mu \nu}
	=
	2(e J)^2
	\sum_{\lambda,\lambda'}
	\int \frac{d^2 k}{(2 \pi)^2}
	V^\mu_\lambda 
	\,
	V^\nu_{\lambda'}
	\,
	\sin\left(\vex{k}\cdot\vex{V}_\lambda\right)
	\,
	\sin\left(\vex{k}\cdot\vex{V}_{\lambda'}\right)
	\,
	n'(\varepsilon^{(\text{SL})}_{\vex{k}}),
\end{align}
where $n'(\e) = (d / d \e) \, n(\e)$. 
Eq.~(\ref{London}) becomes 
\begin{align}\label{London-2}
	\left\langle J^\mu \right\rangle
	=&\,
	-
	\frac{2 e^2 J}{c}
	\int \frac{d^2 k}{(2 \pi)^2}
\left\{
	\sum_{\lambda}
	V^\mu_\lambda 
	\,
	\sin\left(\vex{k}\cdot\vex{V}_\lambda\right)
	\,
	\left[
	\frac{\partial}{\partial k^\nu}
	\,
	n(\varepsilon^{(\text{SL})}_{\vex{k}})
	\right]
	+
	\sum_{\lambda}
	V_\lambda^\mu
	\,
	V_\lambda^\nu
	\,
	\cos(\vex{k}\cdot\vex{V}_\lambda)
	\,
	n(\varepsilon^{(\text{SL})}_{\vex{k}})
\right\}
	\,
	A^\nu
	=
	0.
\end{align}
The last equality follows after integrating by parts.
\end{widetext}
Our result is different from that obtained previously in Ref.~\cite{Yuzbashyan2006Dynamical_Vanishing},
which found that the superfluid density is half that of the superconducting
ground state. 
We have verified our result that the superfluid density vanishes in phase I 
for the continuum particle-hole symmetric model as well. 
Moreover, we have also computed the superfluid density in phase II; we omit details here. 
We find that the superfluid density deviates from the ground state value 
(equal to the total electron density) 
for any nonzero quench in phase II, 
and goes to zero as the quench strength approaches the dynamical phase II-phase I boundary. 
These results are completely consistent with our findings for the Mattis-Bardeen (dirty limit) 
optical conductivity in phases II and I.


\begin{thebibliography}{64}
\bibitem [{qua()}]{quarter_filling}%
\BibitemOpen
\href@noop {} {}\bibinfo {note} {At half filling, the density of states of a
	square lattice contains a van Hove singularity. We choose quarter filling to
	avoid such complication. The results should apply to all filling without
	singular density of states contributions.}\BibitemShut {Stop}%
\end{thebibliography}
\end{document}